\documentclass{ws-ijmpa}
\usepackage[compress]{cite}
\usepackage{graphicx}
\usepackage{breakurl}

\usepackage{epsfig,graphicx}
\usepackage{amssymb,amsmath,amsfonts}
\usepackage{multirow}
\usepackage[dvipsnames]{xcolor}
\usepackage[bookmarks=false]{hyperref}
\usepackage{slashed}

\graphicspath{{figures/}}

\newcommand{\Tr}{\mathop{\mathrm{Tr}}\nolimits}

\newcommand{\psibar}{\ensuremath{\bar\psi}}
\newcommand{\pbp}{\ensuremath{\psibar\psi}}

\catchline{}{}{}{}{}
\begin{document}
\title{Topology and axions in QCD}

\author{Maria Paola Lombardo}

\address{INFN Firenze\\lombardo@fi.infn.it}

\author{Anton Trunin}

\address{Samara U.\\
amtrnn@gmail.com}

\maketitle

\begin{history}
\received{Day Month Year}
\revised{Day Month Year}
\end{history}

\begin{abstract}
QCD axions are at the crossroads of QCD topology and Dark Matter searches. We present here the current status of topological studies on the lattice, and their implication on axion physics. We outline the specific challenges posed by lattice topology,  the different proposals for handling them, the observable effects 
of  topology on  the QCD spectrum and  its interrelation with chiral and axial symmetries.
We  review the transition to the Quark Gluon Plasma,
 the fate of topology at the transition, and the approach to the  high temperature limit. We discuss the extrapolations needed to  reach the regime of cosmological relevance, and the resulting constraints on the QCD axion. 
\end{abstract}


\tableofcontents
\section{Overview}

We know that strong interactions have many facets: among the most
fascinating aspects there is the possibility of including a topological
term into the QCD Lagrangian, which leads naturally to the prediction of 
a yet-unobserved particle -- the QCD axion~\cite{Peccei:1977hh,Peccei:1977ur,Weinberg:1977ma,Wilczek:1977pj} -- a theoretically well-motivated candidate for Dark Matter. 
The same topological term solves 
an apparent mystery of the hadron spectrum, giving a mass to the $\eta'$ meson~\cite{tHooft:1976rip,tHooft:1976snw,Weinberg:1975ui}. 

Central in the discussions of this note 
is the relation between the axion mass $m_A$ and the axion decay constant $f_A$, which from the most recent estimates~\cite{diCortona:2015ldu,Gorghetto:2018ocs} is
\begin{equation}
m_A = \frac{\sqrt{ \chi_{top}}}{f_A}= 56.9(5)\, \frac{10^{11} \text{GeV}}{f_A}~\text{$\mu$eV},
\label{eq:basic}
\end{equation}
$\chi_{top}$ is the topological susceptibility of QCD, and the relation above
is valid providing that $f_A$ is much larger than the QCD scale, $f_A \gg \Lambda_{QCD}$.
 
Astrophysical observations
give the approximate limits $ 10^{12} \gtrsim  f_A \gtrsim 4 \times 10^{8}$~GeV;
the upper bound prevents producing an amount of Dark Matter exceeding its estimated contribution, the lower one limits the amount of energy from the observed neutrino cooling of supernova 
SN1987A \cite{Mayle:1987as,Raffelt:2006cw,Vysotsky:1978dc, Braaten:2019knj,Ringwald:2018xlf}.

A range of decay constants $f_A$ exists for which the QCD 
axion would be a possible cold Dark Matter candidate\cite{Preskill:1982cy,Abbott:1982af,Dine:1982ah,Wantz:2009it,Marsh:2015xka,Rubakov:2019lyf,Ballesteros:2019tvf}.
This range  has to be estimated by considering
the axion's cosmological history: 
in these analyses, the temporal evolution~--
or, equivalently, the temperature dependence~--
of the topological susceptibility in QCD plays an important role.

QCD can be studied in the framework of perturbation theory, and many important results
have been obtained within this approach. However, it turns out that topology
is completely outside the domain of perturbation theory: within a perturbative
approach the contribution of the topological term would always be  zero. 
Topology in QCD is related to the mechanisms of chiral and axial
symmetries breaking and restoration. Chiral symmetry breaking at
low temperatures, and instanton models combined
with  perturbation theory 
at very high temperatures,
dictate the behaviour of the topological observables in these
limiting situations~\cite{Gross:1980br,Ringwald:1999ze,diCortona:2015ldu,Gorghetto:2018ocs,Bottaro:2020dqh}.  

At a temperature of about 150 MeV ($T_c = 154(3)$ MeV according to the latest estimates \cite{Ding:2019prx}) chiral symmetry is 
approximatively restored, and 
quarks and gluons are  in their plasma phase  --
the Quark Gluon Plasma. 
Temperatures ranging from $T_c$ till $\approx 2 T_c$  are explored within experimental
studies of Quark Gluon Plasma, higher 
temperatures  -- above 500~MeV -- become relevant for cosmology, with strong interactions still playing a significant
role. No analytic 
approach is quantitatively satisfactory
for temperatures ranging from $T_c$  till $T \approx 2$~GeV. 
Within this range, ab-initio lattice simulations are mandatory. 

High temperature lattice studies~\cite{DElia:2018fjp,Philipsen:2019rjq,Ding:2020rtq} face specific challenges: 
around the transition to Quark Gluon Plasma one has to deal with a pseudo-critical dynamics.
At higher temperatures there are other challenges:  
lattice simulations are done on a discretized Euclidean space
 where the inverse of the (compactified) time direction $a N_t$
 gives the temperature $T = 1/ a N_t$.
With the current lattice spacings $a \simeq 0.05$~fm
such high temperatures would mean three--four points in time
directions: the space direction would get correspondingly small, 
while low lying modes would push the correlation lengths to large values, making finite size effects
particularly dangerous. In addition to this, 
lattice topology~\cite{Muller-Preussker:2015daa,Bonati:2017nhe} poses specific problems.
We will devote one Section to the discussions of QCD topology
on the lattice and of the various methods that have been employed to address these issues. 
The final two Sections will present the current status of the results, with special
attention to recent lattice works that for the first time addressed QCD axion physics
from first principles\cite{Berkowitz:2015aua,Borsanyi:2016ksw,Petreczky:2016vrs,Bonati:2015vqz,Burger:2017xkz,Burger:2018fvb}.

In this review  we will only discuss the QCD axion. Let us just mention that
axions' physics is a much broader topic:  axions appear in many different
extensions of the Standard Model to explain the lack of observed CP violation
of strong interactions. In string theories axions are ubiquitous, as 
they are generated by their complex topology~\cite{Svrcek:2006yi}. 
In most (or all)  models considered so far, axions are fundamental scalar 
fields, however recently models for composite axions have been considered as well~\cite{Anastasopoulos:2018uyu,Bigazzi:2019eks}. Axions are subjected to extensive experimental 
searches: experiments are optimized for certain mass ranges, and constraints
from theories play an important role. For these important topics, which are
not however in the scope of our discussion, we
refer the reader to  recent  works and comprehensive reviews~\cite{Braaten:2019knj,DiVecchia:2019ejf,Rubakov:2019lyf,Irastorza:2018dyq,Sikivie:2020zpn,DiLuzio:2020wdo,Jaeckel:2020dxj}. 

In brief summary, our discussion of topology and axions 
calls into play  several
entangled topics: QCD phenomenology, hadron spectrum, phases and
phase transitions in QCD, early Universe and particle cosmology.
At the same time, each of the two aspects --
QCD topology and axions -- has independent reasons of interest. We have thus organised the material trying to outline  self-consistent discussions, with the aim to provide
the basic information and tools to follow the current
literature on QCD topology, and its implications on axion physics.

The material is organised as follows:  in the next Section we will review the strong CP problem 
and introduce the axion. Section~\ref{sec:spectrum} discusses topology
in the context of symmetries of QCD,  the solution to the $U(1)_A$  puzzle and its role
at the QCD transition. Section~\ref{sec:lattice} is devoted to the
main theoretical tool of this review, the lattice formulation. After a brief introduction, the focus
is on the methods for topology. Next, we present 
the lattice results for the topological susceptibility at high temperature, and finally
we discuss the impact of these results on 
axion physics.

\section{Topology and the Strong CP problem}

The QCD Lagrangian admits a CP violating term
\begin{equation}
\mathcal L=  {\cal{L}}_{QCD} + \theta\frac{g^2}{32 \pi^2}  F^a_{\mu \nu}\tilde {F}_a^{\mu \nu},
\label{eq:lqcd}
\end{equation}
where
$\dfrac{g^2}{32 \pi^2} F^a_{\mu \nu} \tilde{F}_a^{\mu \nu}$
is the topological charge density $q(x)$, $\tilde F^{\mu\nu}=\dfrac12\varepsilon^{\mu\nu\rho\sigma}{F}_{\rho\sigma}$,
and $\theta\,q(x)$ is known as the $\theta$-term. 

Without the $\theta$-term strong interactions conserve CP. 
Once this term is included,
the neutron acquires an electric dipole moment $d_n$, which 
can be estimated 
with QCD sum rules:  $d_n = 2.4 \times 10^{- 16} \theta$ e cm
\cite{Pospelov:1999mv} and
chiral perturbation theory,  
$d_n = 3.3 \times 10^{-16} \theta$ e cm \cite{Pich:1991fq}, 
and $d_n = 3.6 \times 10^{-16} \theta$  e cm\cite{Pospelov:1999mv}.  
The most recent experimental measure\cite{Abel:2020gbr} of the neutron
electric dipole moment is 
$d_n=(0.0 \pm 1.1$ (stat) $\pm 0.2$ (sys)) $\times 10^{-26}$ e cm,
which maybe interpreted as an upper limit $|d_n| < 1.8 \times 10^{-26}$ e cm at a $90 \%$ C.L,
leading to the  bound $\theta <0.5 \times  10^{-10}$. While the new
results place more and more stringent limits, the unnaturally small value
of $\theta$ has been known since long.
This is known as the strong CP problem -- we will come back to its solution
in the next Section. 

The  Grand Canonical partition function of QCD is now a 
function of $\theta$. Moreover,
for future discussion, we  make explicit a dependence on the temperature $T$:
\begin{equation}
\label{eq:Z_QCD}
\mathcal Z (\theta, T) = \int \!\mathcal D [\varPhi] \,e^{-T \sum_t \int d^3 x {\cal{L}}(\theta)}\
 = e^{-V F(\theta, T)}.
\end{equation}
Let us consider the $\theta$ dependent energy density $F(\theta, T)$
\cite{Vicari:2008jw}. 
$F(\theta, T)$ is related with the probability of finding 
configurations with given topological charge $Q=\int d^4x\, q(x)$:
\begin{equation}
P_Q = \int\limits_{-\pi}^{\pi} \frac{d \theta}{2 \pi} 
e^{-i \theta Q} e^{- V F(\theta)},
\end{equation}
so the coefficients $C_n$ of the Taylor expansion 
\begin{equation}
F(\theta, T) = \sum_{n=1}^{\infty} (-1)^{n+1} \frac{\theta^{2n}}{2n!}\, C_n
\label{eq:Taylor}
\end{equation}
are given by the cumulants of the topological charge: 
\begin{equation}
C_n = (-1)^{n+1}  \frac{d^{2n}}
{d\theta^{2n}}F(\theta, T)\Bigl|_{\theta = 0} = \langle Q^{2n}\rangle_{conn}.
\label{eq:cumulants}
\end{equation}
It can be shown that the free energy  $F(\theta, T)$
as a function of $\theta$ has a minimum at $\theta=0$ \cite{Kim:1986ax}: however,
this per se does not solve the strong CP problem, as
 at this stage  $\theta$ is just a parameter.

It is convenient to introduce 
the dimensionless scaling function $f(\theta)$:
\begin{equation}
f(\theta,T) = (F (\theta, T) - F (0, T))/M^4,
\end{equation}
where $M$ is any suitable energy scale,
for instance $\chi_{top}^{1/4}$.
By comparing this with the
Taylor expansion~\eqref{eq:Taylor}, it is natural to parametrize $f(\theta)$ as
\begin{equation}
f(\theta, T) = \frac12 C\,\theta^2 s(\theta,T),
\end{equation}
where $s$ is a dimensionless function, and $s(\theta\!=\!0) = 1$.
Then 
\begin{equation}
s(\theta, T) = 1 + b_2(T)  \theta^2 + b_4(T) \theta ^4 +\ldots,
\label{eq:bs}
\end{equation}
and the $b$'s are easily expressed in terms of the cumulants
$C_n$ in Eq.~\eqref{eq:cumulants}.

At zero and low (below or around $T_c$) temperatures 
$F(\theta,T)$ can be computed by considering 
low energy effective Lagrangians \cite{DiVecchia:1980yfw,Rosenzweig:1979ay}
-- these are actually the calculations
leading to the estimate of the neutron magnetic dipole moment
mentioned above.  A recent study improved 
the precision   by including electromagnetic corrections
and NNLO corrections in the chiral expansion \cite{Gorghetto:2018ocs}.
The topological susceptibility and  the fourth-order cumulant 
have been computed up to NNLO in $U(3)$ Chiral Perturbation 
Theory\cite{Nicola:2019ohb} at zero and non-zero temperature, as well as within Interacting Instanton Liquid Models \cite{Schaefer:2013isa,Wantz:2009it}.
Finite temperature corrections
have been computed 
\cite{diCortona:2015ldu,Bottaro:2020dqh}, and their validity stretches till $T \simeq T_c$. 

In the Quark Gluon Plasma phase the basic expression for
$F(\theta,T)$ was computed long ago\cite{Gross:1980br, Ringwald:1999ze}. 
The dilute gas approximation (DIGA)
and high temperature perturbation theory at leading order
reads\cite{Gross:1980br,Ringwald:1999ze}:
\begin{equation}
F(\theta,T) - F(0, T) \simeq T^{4 - \beta_0}\left({\frac{m_l}{T}}\right)^{N_{f,l}}(1 - \cos\theta),
\label{eq:DIGAF}
\end{equation}
where $\beta_0 = 11 N_c/3 - 2N_f/3$ and $N_{f,l}$ is the number of light flavors.

The fundamental physical fact\cite{Gross:1980br} is that
at sufficiently high temperature only fields with integral topological
charge can contribute to the functional integral, so
the $\theta$ dependence of the free energy at high temperature is dominated
by  instantons~-- ultimately only by the ones with positive or negative unit charge. 
This contrasts with  low temperature, where  the physics mechanisms leading
to a contribution of the topological $\theta$ term to the partition function
can be understood without invoking instantons \cite{Veneziano:1979ec}. The microscopic 
behaviour of the plasma close to phase transition, 
its interpretation in terms of
physical degrees of freedom, including topological structures, 
is of course a subject of active research \cite{Larsen:2019sdi,Larsen:2018crg,Sharma:2019wiv}.

As already mentioned, the 
main tool to analyze strong interactions in non-perturbative regime
is lattice simulations, which rely on a sampling of the phase space
weighted with the $e^{-S}$, where $S$ the Euclidean action. 
In Euclidean space-time the Minkowskian Lagrangian~\eqref{eq:lqcd}
becomes complex for real values of $\theta$: this hampers a direct importance sampling. However, it is possible to
consider a Taylor expansion
around $\theta$ = 0 -- cf. Eqs. \eqref{eq:cumulants} and \eqref{eq:bs}~--  thus monitoring the
approach to one of the limiting expressions for $F(\theta, T)$\cite{Gross:1980br,diCortona:2015ldu,Gorghetto:2018ocs}.
Clearly this approach is satisfactory only at small $\theta$. For instance,
the interesting physics around $\theta = \pi$ \cite{DiVecchia:2017xpu}
would require a different approach. Some attempts in this direction have been
made~\cite{Bonati:2016tvi} by using an imaginary value of~$\theta$, followed
by an analytic continuation to real values. The topological
susceptibility, whose computations
will be described in the following, is  the leading
order contribution to this series.

\subsection{Solution of the strong CP problem:  the axion}

Suppose $\theta$ in Eq.~\eqref{eq:lqcd}
were a dynamical parameter: in such a case, dynamics would force
its value to zero, thus solving the strong CP problem.
In order to achieve this, the existence of an
extra particle~\cite{Peccei:1977hh,Peccei:1977ur,Weinberg:1977ma,Wilczek:1977pj}
was postulated, a pseudo-Goldstone boson of a spontaneously
broken  symmetry known as the Peccei-Quinn (PQ) symmetry,
which couples to the QCD topological charge, with
a coupling suppressed by a scale $f_A$.
The axion field $a(x)= f_A \theta(x)$ is now a space-time dependent $\theta$  parameter.
The axion--QCD Lagrangian  reads
\begin{equation}
\mathcal L=  {\cal{L}}_{QCD} + \theta\frac{g^2}{32 \pi^2}  F^a_{\mu \nu}\tilde{F}_a^{\mu \nu} +\\
{\partial^2_\mu}{ a^2} + \frac{a}{f_A}\frac{g^2}{32 \pi^2}  F^a_{\mu \nu}\tilde {F}_a^{\mu \nu}.
\end{equation}
Moreover, one assumes that the theory enjoys a shift symmetry:
$a \to a + \alpha$. The $\theta$ angle may
be eliminated by a shift, and the $\theta$ dependence has been traded with
a dependence on the
axion field, whose minimum is at zero \cite{Kim:1986ax}:
this solves the strong CP problem. Besides the original 
papers~\cite{Peccei:1977hh,Peccei:1977ur,Weinberg:1977ma,Wilczek:1977pj},  
there are many  reviews discussing
this point detail\cite{Dine:2000cj,Vicari:2008jw}.

$F(\theta, T)$ can now be used to compute the axion mass. 
At leading order in $1/f_A$~-- well justified as $f_A \gtrsim 4 \times {10^8}$~GeV~-- the axion
can be treated as an external source, and its mass is given by
\begin{equation}
m_A^2(T) f_A^2 = \frac{\partial^2 F(\theta, T)}{\partial \theta^2}\biggl|_{\theta=0} \equiv
\chi_{top}(T) \; .
\end{equation}
At low temperature,  chiral perturbation theory gives the result~\cite{Weinberg:1977ma}
\begin{equation}
m_A^2 = \frac{m_u m_d}{(m_u + m_d)^2} \frac{m_\pi^2 f_\pi^2}{f_A^2},
\end{equation}
which has been recently improved to NLO \cite{diCortona:2015ldu}.
Finite temperature corrections have been
computed\cite{diCortona:2015ldu, Gorghetto:2018ocs} --
the analysis is the same as the one discussed above for QCD topology --
and their validity stretches till $T \simeq T_c$.

At low temperature the LO chiral perturbation theory relation 
between topological susceptibility and chiral condensate \cite{DiVecchia:2017xpu}
ensures that the two expressions coincide. 
More generally, we have now a prescription for
the temperature dependence of the QCD axion mass. We underscore that 
the axion is massive because
the $U(1)$ PQ symmetry is anomalous: 
because of that, the would-be Goldstone is a massive
pseudo-Goldstone scalar. The amount of breaking is regulated by the 
topological susceptibility, hence it is temperature dependent, 
and there is a close relation with chiral symmetry. 

In very brief summary, 
the essence of this discussion
is the close relation between  axion mass and  topological susceptibility: 
\begin{equation}
m_A^2 f_A^2 = \chi_{top},
\label{eq:ritop}
\end{equation}
which is valid for any temperature. 
Inserting the known value of today's topological susceptibility,
we obtain Eq.~(\ref{eq:basic}). 

\section{Topology, symmetries and spectrum of strong interactions}
\label{sec:spectrum}
We have seen that a relation emerges,
which links topological susceptibility and chiral condensate, and that the axion is massive if the topological
susceptibility is non-zero. We will see that a completely analogous mechanism
solves another puzzle in QCD, the $U(1)_A$ problem. In essence, without
a contribution from the $\theta$-term, all isoscalar mesons should be
lighter that $\sqrt3\, m_\pi$. It turns out that the same anomaly giving
mass to the axion is responsible for solving this problem.

The rich behaviour of the strong interactions is encoded in the 
apparently simple QCD Lagrangian
\begin{equation}
{\cal{L}} = \sum_{a=1}^n  \bar{q}_{La} \slashed \partial {q}_{La} +       
\bar{q}_{Ra} \slashed\partial  {q}_{Ra} -m  (\bar q_{La}q_{La} +
\bar{q}_{Ra} {q}_{Ra}) + \theta\,\frac{g^2}{32\pi^2} 
F^a_{\mu \nu} \tilde F_a^{\mu \nu} + {\cal{L}}_{gauge},
\end{equation}
where $q$ are the quark fields and $q_{L,R} = \dfrac{1 \mp \gamma_5}{2} q$.
In this form, and with $m=0$,  
the invariance under the transformation $q_L \to V_L q_L$,
$q_R \to V_R q_R$, with $V \in U(n)$, is manifest. 
Thus, at classical level there is a global symmetry 
$U(n)_L\!\times\!U(n)_R \cong SU(n)\!\times\!\
SU(n)\!\times\!U(1)_V\!\times\!U(1)_A$.
The pseudoscalar mesons are candidate Goldstone bosons if
the symmetry $SU(n)\!\times\!SU(n)$ is spontaneously broken.
With finite masses we have to consider approximate symmetries, either
$SU(2)\!\times\!SU(2)$, assuming that the strange quark does not contribute to chiral
dynamics, or $SU(3)\!\times\!SU(3)$ including it. 
The pion triplet has masses of about $140$~MeV, the meson containing strange quarks $\eta$ about $540$~MeV, the four $K$'s at about $400$~MeV, and finally the $\eta'$ much heavier, $960$~MeV.
The most natural scenario accommodating
experimental observations of pions and $K$ mesons
is the spontaneous breaking of the
$SU(3)\!\times\!SU(3)$ symmetry. This breaking should be accompanied by the
formation of a quark condensate, which, in turn, would spontaneously break $U(1)_A$.  Hence  $\eta'$ should follow the same fate as the
other mesons, which clearly it is not the case. 
The way out is a breaking of the $U(1)_A$ symmetry\cite{Weinberg:1975ui}: 
the same topological
structures which are responsible for the axion mass give mass to the $\eta'$. 
With the  $q_L \to e^{-i \theta} q_L$, $q_R \to e^{i \theta}q_R$ violation, 
the remaining symmetry is then 
$U(n)_L\!\times\!U(n)_R / U(1) \cong SU(n)\!\times\!SU(n)\!\times\! U(1)_V$. 
The experimental value of the $\eta'$ mass gives an experimental evidence to
the explicit $U(1)_A$ breaking.

Theoretically, the breaking can be explored by lattice simulations.
The topological contribution to $\eta'$ mass is given by the Witten-Veneziano formula~\cite{Witten:1979vv,Veneziano:1979ec,Feldmann:1999uf}:
\begin{equation}
\label{eq:wv-fomula}
\chi_{top}\Bigl|_{N_f=0}=\frac{1}{2N_f}f_0^2\left(m_{\eta'}^2+m_{\eta}^2-2m_K^2\right),
\end{equation}
where $\chi_{top}$ has to be calculated in pure Yang-Mills theory corresponding to the quenched limit $N_f=0$ of QCD. Note that in the chiral limit both $\eta$ and $K$ masses in the r.h.s. of Eq.~\eqref{eq:wv-fomula} vanish.
The topological susceptibility is accessible in lattice simulations (see Section~\ref{sec:lattice} for details), and the relation~\eqref{eq:wv-fomula} was confirmed on the lattice with high accuracy~\cite{DelDebbio:2004ns,Durr:2006ky,Cichy:2015jra}.
So, it is safe to declare that the solution to the $U(1)_A$ problem has been
confirmed also by these ab initio studies.
The natural question then arises, concerning the interrelation
of chiral and axial symmetries at finite temperature, 
and the fate of the $\eta'$, a central
issue of strong interaction physics which is addressed by
model studies~\cite{Nicola:2019ohb,Horvatic:2018ztu},
phenomenological analysis \cite{Kapusta:2019ktm,Kapusta:1995ww,Shuryak:1993ee}, 
FRG studies~\cite{Resch:2017vjs,Schaefer:2013isa,Gao:2020qsj}, and mostly
by lattice analysis~\cite{Mazur:2018pjw,Sharma:2018syt,Fukaya:2017wfq,Sharma:2017yjc,Schmidt:2017bjt,Tomiya:2016jwr,Kotov:2019dby,Kotov:2020hzm}. 

The fate of $U(1)_A$ has implications on the nature of the chiral
phase transition. If the axial symmetry breaking is not much sensitive
to the chiral restoration, the breaking pattern is indeed 
$SU(2)_L\!\times\!SU(2)_R \to SU(2)_V$  or
$O(4) \to O(3)$ symmetry\cite{Pisarski:1983ms}. In this case the universality class 
is well known,
and would correspond to a second order transition with
known exponents and equation of state.  Alternatively,
if the axial symmetry is effectively restored at the same temperature (it cannot be
restored before) as the chiral transition, 
then the relevant symmetry would be 
$U(2)_L\!\times\!U(2)_R \to U(2)_V$, which would hint either at a first 
or
even a second order transition with different exponents \cite{Pelissetto:2013hqa}. 

\begin{figure}
\includegraphics{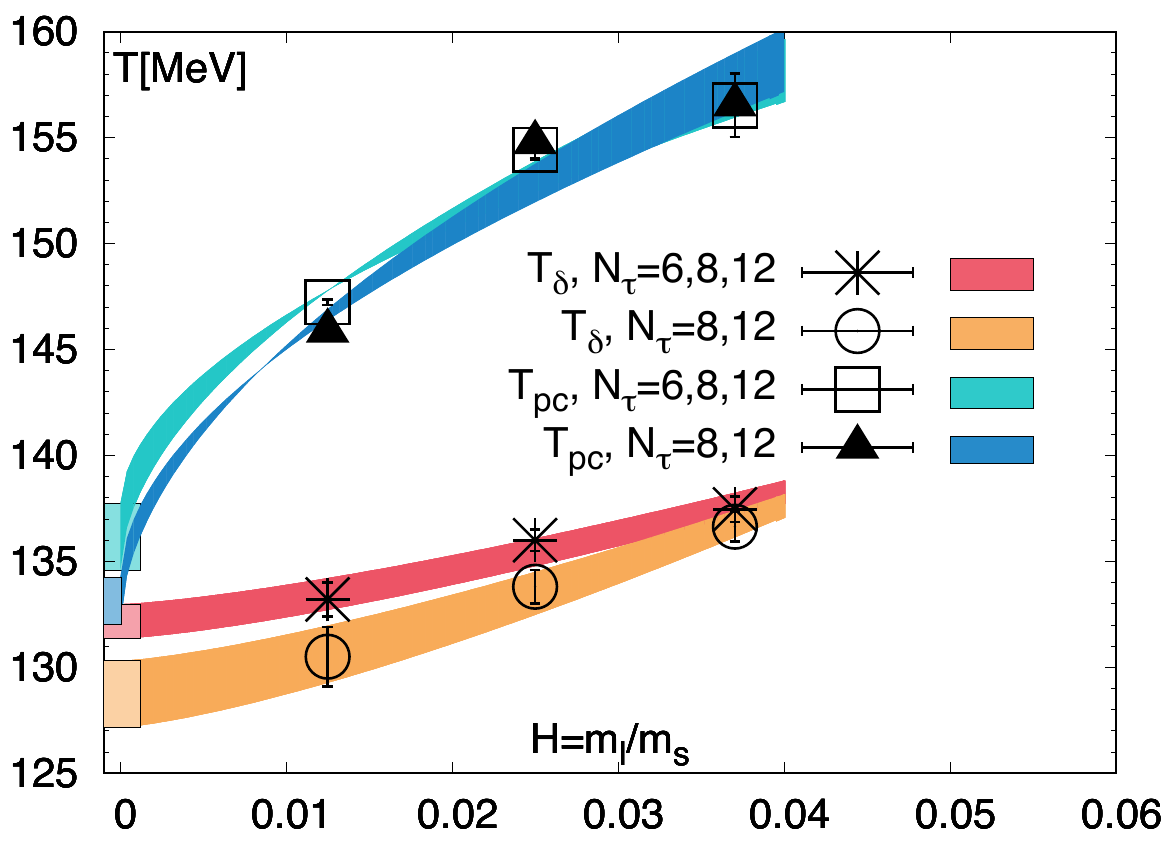}
\caption{The pseudo-critical temperatures from HotQCD from different
observables, and their extrapolation to the chiral limit from
a recent review \cite{Ding:2020rtq} (see also\cite{Ding:2019prx}).}
\label{fig:tcvsmq}
\end{figure}

Lattice simulations can be performed with varying values of  quark masses. 
When addressing the
issue of the universality class of the transition one takes advantage of this 
possibility  to study the associate mass
dependence:  different universality classes have distinct predictions for the 
mass dependence  of the transition temperature.
In principle, the universality class of the transition, as well as the critical temperature
in the chiral limit, could be inferred
by contrasting  lattice results with theoretical predictions
\cite{Philipsen:2019rjq,DElia:2018fjp,Ding:2019prx}.
One typical choice is the pseudocritical temperature as a function of
the quark mass $T_c(m_\pi)$, which within the scaling window
of the theory should follow
\begin{equation}
T_c(m_\pi) = T_c(0) + A\, m_\pi ^{2/{\beta \delta}}.
\end{equation}
In practice, the exponents characterizing different critical behaviours
change very little: $2/\beta \delta \approx 1.08$ for $O(4)$ and 
$2/{\beta \delta} \approx 1.28$ for a $Z_2$ universality
class associated with a hypothetical endpoint of a first order transition
in the chiral limit\cite{Burger:2011zc}. 
The current status of the QCD transition is shown
in  Figure~\ref{fig:tcvsmq}. The heavier
quark mass is at its physical value, and we read off from the plot the accepted value of the pseudo-critical temperature of the QCD crossover: $T_c = 154.3(2)$~MeV.
The fit to the data is  consistent with an $O(4)$ transition, but others cannot be ruled out\cite{Ding:2019prx,Philipsen:2019rjq}.

Results from Wilson fermions exist as well, and 
confirm this picture\cite{Kotov:2019dby,Kotov:2020hzm,Aarts:2019hrg}. Besides giving information on the possible universality classes, and on the physical transition,  Figure \ref{fig:tcvsmq} shows the critical temperature in the chiral limit -- a genuine singular point of strong interactions.
This temperature marks the beginning of the Quark Gluon Plasma and the high temperature phase of QCD. 

More direct approaches to the analysis of
the breaking and restoration of chiral
symmetry based on the consideration of (approximate) order parameters
or Dirac spectrum have not been conclusive
regarding the fate of the $U(1)_A$ symmetry. 
The main observable characterizing chiral symmetry is the chiral condensate,
the order parameter of the $SU(2) \! \times \! SU(2)$ symmetry. When analyzing
axial symmetry, it is useful to consider a fuller set of chiral observables
associated to vector and axial symmetries: 
one can consider the correlation functions of local operators\cite{Shuryak:1993ee,Buchoff:2013nra}
{\allowdisplaybreaks
\begin{eqnarray}
  \sigma &=& \bar{\psi}_l \psi_l  \notag \\
  \delta^i &=& \bar{\psi}_l \tau^i \psi_l \\
  \eta      &=& i\bar{\psi}_l \gamma^5 \psi_l \notag\\
  \pi^i      &=& i\bar{\psi}_l \tau^i \gamma^5 \psi_l. \notag
\label{eq:mesons}
\end{eqnarray}}$(\sigma, \pi^i)$ and $(\eta, \delta^i)$ are related by $SU(2)_L \!\times\! SU(2)_R$
transformations, $(\sigma, \eta)$, $(\delta^i, \pi^i)_{1\le i \le 3}$ via
$U(1)_A$. Inspecting the degeneracy, or lack thereof, of these two point functions,
or, equivalently, of the associated susceptibilities (integrals over the four
space) one accesses information on the realization  of symmetries. First results including the $\eta'$ have appeared recently\cite{Kotov:2019dby,Kotov:2020hzm}. The
analysis is complicated by the explicit breaking induced by the quark mass,
by the ultraviolet divergences of the susceptibilities, and by the possible
occurrence of accidental degeneracies \cite{Cohen:1996ng}.
Despite these difficulties, 
there is consensus
that axial symmetry is effectively restored  above $T_c$,
say $T \simeq 1.2 T_c$ \cite{Kaczmarek:2020sif,Mazur:2018pjw,Buchoff:2013nra,Suzuki:2020rla,Kanazawa:2015xna,Aoki:2012yj,Tomiya:2016jwr,Brandt:2019ksy,Brandt:2016daq,Cossu:2013uua,Chiu:2013wwa,Tomiya:2016jwr},
however it is unclear how close to $T_c$ the effective
 restoration may happen. There are also attempts to 
tackle this issue within
a first-principle FRG approach \cite{Braun:2020ada}, but at 
the moment
the interrelation of  chiral and
$U(1)_A$ restoration remains an unresolved problem of QCD.

As already stressed,  the anomaly effects breaking the $U(1)_A$ 
symmetry are related to the topological properties of QCD,
so these uncertainties further add to the interest of topological
studies. At a more practical level, the restoration of the $SU(2) \times SU(2)$  opens the way to a simple measurements of the topological susceptibility at high temperatures. We will return to this point in the next Section. 

\section{Lattice Field Theory: methods and results}
\label{sec:lattice}
\subsection{Lattice QCD: brief introduction}
The essentially nonperturbative nature of topology and related issues renders the usual perturbative methods of quantum field theory to be of limited use, so alternative approaches are required. Currently, one of the most prospective and rewarding approaches is lattice formulation of QCD, which allows first-principle calculation of topological and other physical quantities in numeric simulations. In Lattice QCD the continuous spacetime is replaced by discrete 4D Euclidean lattice, and so discretized versions of operators defined on sites of the lattice are introduced. Then, expectation value of observable $O$ can be defined as
\begin{equation}
\label{eq:O-val}
\langle O\rangle = \frac1Z\int \!\mathcal D[U] \,e^{-S_{lat}[U]}O[U],
\end{equation}
where
\begin{equation}
\label{eq:Z-lat}
Z=\int \!\mathcal D[U] \,e^{-S_{lat}[U]}
\end{equation}
is discretized partition function~\eqref{eq:Z_QCD}, and $S_{lat}$ is lattice action based on the continuum QCD Lagrangian. One of the key parameters is number of lattice points in space (time) directions $N_s$($N_t$) and lattice spacing $a$ between two consequent points. In modern simulations, the values $N_s=128$ and $N_t=256$, or even higher, are reached, with lattice spacings of order $a\sim10^{-2}-10^{-1}$~fm. 

The integrals in Eqs.~\eqref{eq:O-val}--\eqref{eq:Z-lat} are taken over all lattice gauge variables $U$, totaling up to $\mathcal O(N_s^3N_t)$ integration variables. With direct evaluation being obviously impossible, a numerical approximation is applied:
\begin{equation}
\label{eq:O-val-approx}
\langle O\rangle \approx \frac1N\sum_i O[U^{(i)}],
\end{equation}
where the set of lattice configurations $U^{(i)}$, $i=1,\ldots,N$ are generated by means of Monte Carlo algorithm to satisfy probability distribution $\propto e^{-S_{lat}[U]}$. Note that the error of approximation~\eqref{eq:O-val-approx} decreases with the number of generated configurations $N$ as $\mathcal O(1/\sqrt N)$.

From practical point of view, the main drawback of the outlined approach is that the realistic simulations, especially at physical quark masses, consume considerable computational resources, so vast allocations at, in general, supercomputer facilities are required. Also, direct lattice simulations are not possible in some physically interesting cases, such as QCD with finite chemical potential or non-zero $\theta$-term, since the weighting factor $e^{-S_{lat}[U]}$ becomes complex-valued and cannot serve as probability measure anymore.

From theoretical point of view, the way to discretize QCD operators on the lattice, including lattice action $S_{lat}$ itself, is not unique. This leads to several possible options, each with their own advantages and drawbacks, with the arising discretization artifacts (also called lattice artifacts due to their origin in finiteness of lattice spacing $a$) to be dealt in controllable way. In general, it is feasible to have several lattice calculations based on different choices of lattice action, cross-checking and compensating each other, in order to have solid and well-understood physical results. The constant progress in theoretical methods, simulation algorithms and hardware efficiency has already made such cross-checks possible, with physical $N_f=2+1+1$ quark masses reached in several independent studies~\cite{Bhattacharya:2014ara, Borsanyi:2016ksw,Bazavov:2018mes,Alexandrou:2018egz}. 
For more details on lattice field theory, fermionic discretizations and corresponding subtleties we refer to comprehensive presentations in Refs.~\cite{Rothe:1992nt,Montvay:1994cy,DeGrand:2006zz,Gattringer:2010zz,Lellouch:2011zz}.
\subsection{Topological charge on the lattice}
Topology on the lattice can be measured by several methods. Let us start from the so-called gluonic definition, which is based on continuum definition for topological charge density in pure gluonic Yang-Mills theory\footnote{Here, in comparison with Eq.~\eqref{eq:lqcd}, we incorporate the coupling constant~$g$ directly into the gauge fields, as it is customary for lattice formalism.}:
\begin{equation}
\label{eq:q-dens}
q(x)=\frac1{32\pi^2}\varepsilon_{\mu\nu\rho\sigma}\Tr[F^{\mu\nu}(x) F^{\rho\sigma}(x)],
\end{equation}
where $F^{\mu\nu}(x)$ is continuous field strength tensor. Then, the topological charge is defined as
\begin{equation}
\label{eq:top-charge}
Q=\int\! q(x)\,d^4x.
\end{equation}
It can be shown (see, e.g., \cite{Jackiw:2008}), that $Q$ equals to the so-called Pontryagin index or winding number of gauge fields, which can only assume integer values.

The definitions~\eqref{eq:q-dens}--\eqref{eq:top-charge} are straightforwardly implemented on the lattice:
\begin{equation}
\label{eq:Q-lattice}
Q=\frac{a^4}{32\pi^2}\varepsilon_{\mu\nu\rho\sigma}\sum_n\Tr[F_{lat}^{\mu\nu}(n) F_{lat}^{\rho\sigma}(n)],
\end{equation}
where summation is over all sites of the lattice. The discretized field strength tensor can be simply evaluated as traceless antihermitian part of the elementary plaquette $U_{\mu\nu}$, $F_{lat}^{\mu\nu}\propto U_{\mu\nu}\bigl|_{\footnotesize\begin{smallmatrix}\text{\!\!\!traceless}\\ \text{antiherm.}\end{smallmatrix}}\equiv(U_{\mu\nu}- U^\dagger_{\mu\nu})-\frac13\Tr(U_{\mu\nu}- U^\dagger_{\mu\nu})$, and vast majority of improvements with respect to the terms of higher order in lattice spacing $a$ are possible~\cite{BilsonThompson:2001ca,BilsonThompson:2002jk}.
The plaquette $U_{\mu\nu}$ represents product of four oriented gauge variables forming closed loop on the lattice:
\begin{equation}
\label{eq:plaquette}
U_{\mu\nu}(n)=U_{\mu}(n)U_{\nu}(n+\hat\mu)U^\dagger_{\mu}(n+\hat\nu)U^\dagger_{\nu}(n),
\end{equation}
where $U_{\mu}(n)$ denotes gauge variable at the site $n$ pointing along the $\hat\mu$ axis.

Still, it turns out that even with highly improved definitions of $F_{lat}^{\mu\nu}$, the topological charge calculated on the lattice from Eq.~\eqref{eq:Q-lattice} is non-integer. The reason, along with the lattice artifacts, is ultraviolet fluctuations of gauge variables, which have to be additionally renormalized. The alternative is to "smooth" UV fluctuations in gauge fields prior to applying definition~\eqref{eq:Q-lattice} directly, for which many methods have been used: cooling and over-improved cooling~\cite{BilsonThompson:2001ca}, different smearing techniques~\cite{Hasenfratz:2001hp,Morningstar:2003gk,Moran:2008ra} and other smoothing algorithms~\cite{Muller-Preussker:2015daa}. 
One of the recent approaches is smoothing by gradient flow~\cite{Luscher:2010iy}, which has prominent advantage of its renormalization properties proven at all orders in perturbation theory~\cite{Luscher:2011bx}, making it theoretically better established than other methods.

The gradient flow is applied to gauge variables by differential equation
\begin{equation}
\label{eq:gf}
\dot V_\mu(n,\tau) = -g^2[\partial_{n,\mu} S_G(V(\tau))]V_\mu(n,\tau),\qquad
V_\mu(n,0)=U_{\mu}(n),
\end{equation}
where flow time $\tau$ determines the amount of smoothing. Gradient flow can be based on arbitrary choice of gauge action~$S_G$ in~\eqref{eq:gf}, although the simplest Wilson gauge action is mostly used. On practice, Eq.~\eqref{eq:gf} is solved by standard numerical methods for differential equations, such as Runge--Kutta scheme~\cite{Luscher:2010iy}.

\begin{figure}[t!]
\begin{center}
\includegraphics[scale=1.3]{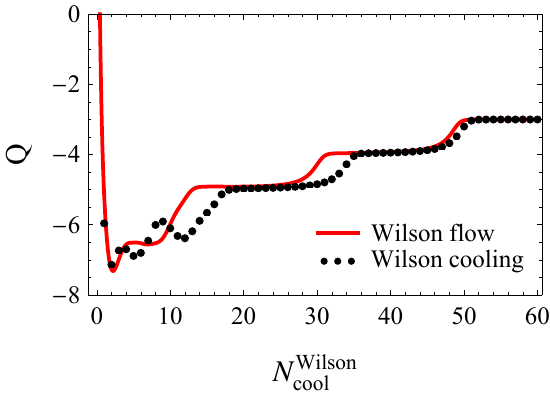}
\end{center}
\caption{Wilson cooling and gradient flow histories for topological charge of a single lattice configuration. $N^\text{Wilson}_\text{cool}$ denotes number of cooling steps.}
\label{fig:single_conf}
\end{figure}

\begin{figure}[t!]
\begin{center}
\includegraphics[scale=1.3]{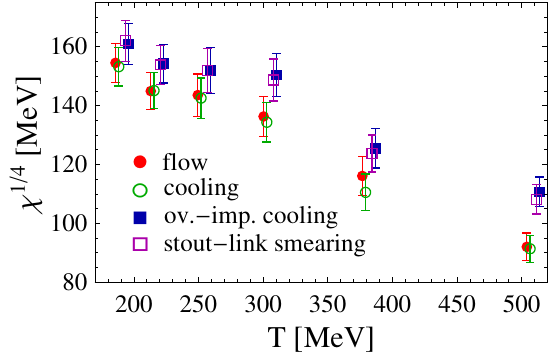}
\end{center}
\caption{Topological susceptibility $\chi_{top}\propto Q^2$ calculated by four different smoothing methods at finite temperature~\cite{Trunin:2015yda}.}
\label{fig:chitop_4cmp}
\end{figure}

It was shown~\cite{Bonati:2014tqa,Alexandrou:2015yba,Alexandrou:2017hqw}, that in leading order of perturbation theory cooling, smearing and gradient flow lead to the identical result for an updated gauge variable, allowing to derive a relation between flow time~$\tau$ and number of cooling/smearing steps. Moreover, numerical measurements revealed that even with relatively large amount of smoothing applied, all considered methods give highly correlated and, in most cases, almost identical results in both zero and finite temperature studies~\cite{Bonati:2014tqa,Namekawa:2015wua,Alexandrou:2015yba,Alexandrou:2017hqw,Cichy:2014qta,Trunin:2015yda}.
For illustration, we show in Figure~\ref{fig:single_conf} comparison of typical Wilson cooling and gradient flow histories for topological charge of a single lattice configuration. In Figure~\ref{fig:chitop_4cmp} we calculate the topological susceptibility 
\begin{equation}
\label{eq:chi-def}
\chi_{top}=\frac{\langle Q^2\rangle}{V}, \qquad V=a^4N_s^3N_t
\end{equation}
by averaging squared topological charge $Q^2$ on the sets of configurations at different temperatures and compare four methods: Wilson flow and cooling with over-improved stout-link smearing and cooling~\cite{Trunin:2015yda}.

Another approach to topology on the lattice is established by Atiyah-Singer index theorem~\cite{Atiyah:1971rm,Atiyah:1984tf}
\begin{equation}
\label{eq:at-theorem}
\frac1{32\pi^2}\,\varepsilon_{\mu\nu\rho\sigma}\!\int \Tr[F^{\mu\nu}(x) F^{\rho\sigma}(x)]\, d^4x=n_+-n_-
\end{equation}
relating topological charge to the number of zero modes of the massless Dirac operator with positive and negative chirality $n_\pm$.
Eq.~\eqref{eq:at-theorem} remarkably connects a purely gluonic quantity with the properties of fermionic operator, so it is called the fermionic definition of topological charge. A principal possibility to implement the Dirac operator satisfying theorem~\eqref{eq:at-theorem} on the lattice was proven in~\cite{Luscher:1998pqa}, and Neuberger's implementation of overlap Dirac operator~\cite{Neuberger:1997fp,Neuberger:1998wv} is most commonly used on practice.

The fermionic definition~\eqref{eq:at-theorem} is appealing in many ways: it has a solid theoretical basis, not affected but UV fluctuations (no prior smoothing of gauge fields needed) and provides integer values of $Q$ by design.
The main drawback, however, is very high computational cost of calculating low-lying modes for any implementation of massless Dirac operator on the lattice.
As an alternative, one can consider Wilson--Dirac spectral flow~\cite{Edwards:1998sh} and stochastic spectral projector~\cite{Giusti:2008vb,Luscher:2010ik} approaches, which are still closely related to the definition~\eqref{eq:at-theorem}.

Finally, a simple (although, only approximate) way to express the topological susceptibility~\eqref{eq:chi-def} through fermionic observable  was outlined in Refs.~\cite{Kogut:1998rh,Bazavov:2012qja,Buchoff:2013nra} on the basis of QCD symmetry relations.
In the continuum theory,
\begin{equation}
\label{eq:topqfer}
\frac1{32\pi^2}\,\varepsilon_{\mu\nu\rho\sigma}\!\int \Tr[F^{\mu\nu}(x) F^{\rho\sigma}(x)]\, d^4x = m_l \int d^4 x\, \bar \psi(x) \gamma_5 \psi(x),
\end{equation}
where $m_l$ is the light quark mass. By squaring Eq.~\eqref{eq:topqfer} and averaging over gauge fields, we immediately get
\begin{equation}
\label{eq:chit-chi5}
\chi_{top}=\frac{\langle Q^2\rangle}{V}=m_l^2\,\chi_{5,disc}.
\end{equation}
The disconnected pseudo-scalar susceptibility $\chi_{5,disc}$ is known to suffer from large fluctuations, so its direct calculation is cumbersome.
Instead, we can utilize the fact that with restoration of chiral symmetry $\chi_{5,disc}$ becomes equal to the disconnected chiral susceptibility~$\chi_{disc}$.
Then, we finally arrive to
\begin{equation}
\label{eq:chit-pbp}
\chi_{top}(T\gtrsim T_c)=m_l^2\,\chi_{disc}=m_l^2\,\frac{V}{T}\left( \langle{(\bar\psi \psi)^2}\rangle_l - \langle{\bar\psi \psi}\rangle_l^2 \right).
\end{equation}
On the lattice, discretization artifacts have to be taken into account, so Eqs.~\eqref{eq:topqfer}--\eqref{eq:chit-pbp} hold only approximately.
Moreover, Eq.~\eqref{eq:chit-pbp} allows to calculate topological susceptibility only in chirally-symmetric phase, i.e. at sufficiently high temperatures. Still, it turned out to be quite useful~\cite{Petreczky:2016vrs,Burger:2018fvb}, since the chiral condensate $\langle{\bar\psi \psi}\rangle_l$ is easily accessible in lattice calculations, and high-temperature behaviour of $\chi_{top}$ is of great importance to axion physics, as we discuss in details in Section~\ref{sec:axion}.

\begin{figure}[t!]
\begin{center}
\includegraphics[width=\linewidth]{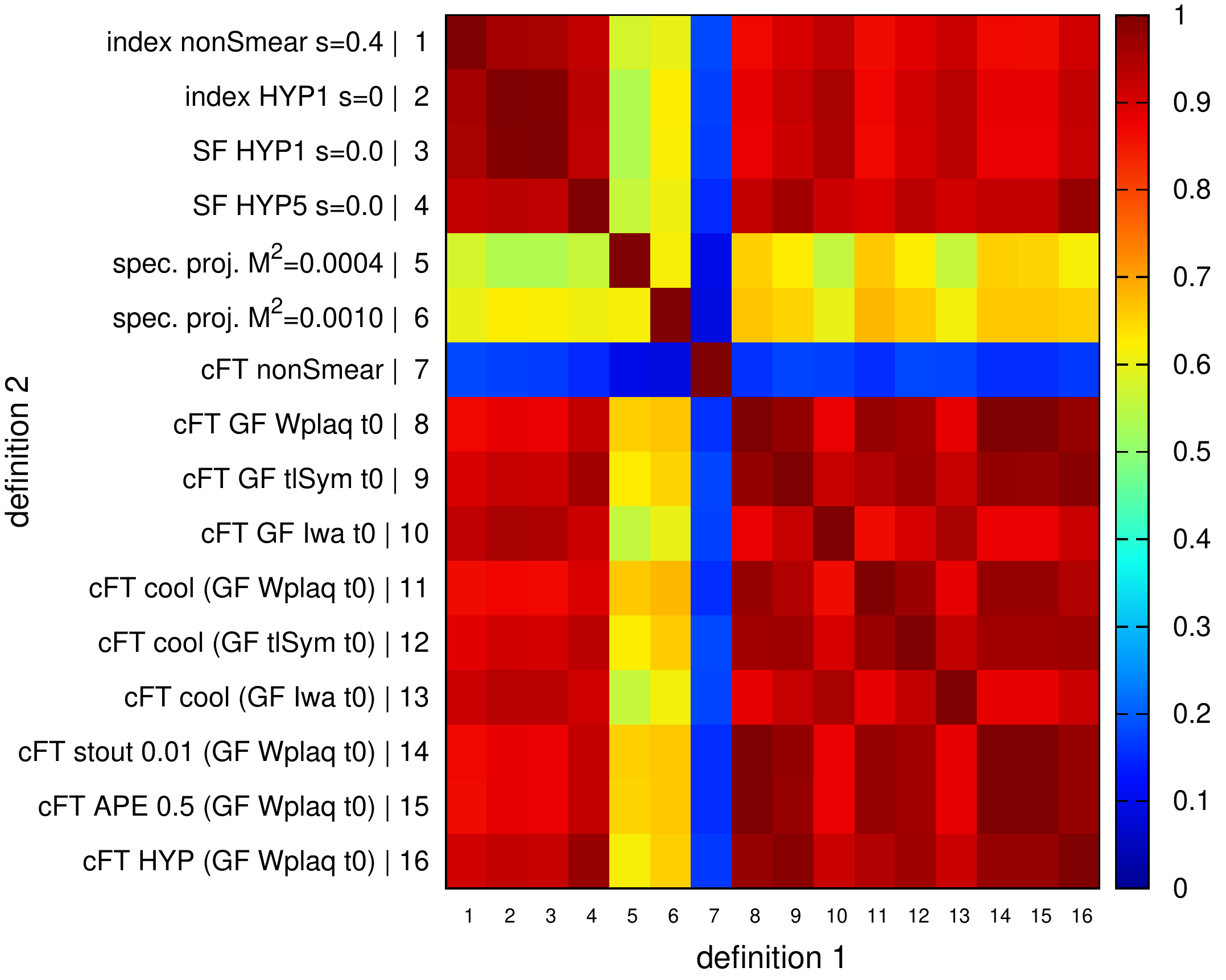}
\end{center}
\caption{Correlation matrix between different gluonic and fermionic definitions of topological charge~\cite{Alexandrou:2017hqw}.}
\label{fig:topcharge_cmp_summary}
\end{figure}

The extensive study and comparison of the different methods for topological charge on the lattice have been carried out in Refs.~\cite{Cichy:2014qta,Alexandrou:2017hqw}. We present in Figure~\ref{fig:topcharge_cmp_summary} the summarizing plot from Ref.~\cite{Alexandrou:2017hqw} containing overall comparison of various gluonic and fermionic definitions outlined above. 
All definitions show from good to perfect agreement between themselves, except for gluonic definition~\eqref{eq:Q-lattice} without any prior gauge field smoothing (\# 7 in Figure~\ref{fig:topcharge_cmp_summary}) and spectral projector results (\# 5 and 6).
As was discussed above, direct measurements with Eq.~\eqref{eq:Q-lattice} are meaningless due to the large UV fluctuations in unsmoothed gauge fields, which is directly confirmed in Figure~\ref{fig:topcharge_cmp_summary}.
Regarding spectral projectors, less correlated with gluonic and even other fermionic methods, we note that for measurements on finite lattices perfect matching between different definitions is not really expected, even for the methods theoretically proven to be identical, due to $\mathcal O(a)$ lattice artifacts.
It was shown in Ref.~\cite{Alexandrou:2017bzk} that in continuum limit $a\to0$ the topological susceptibility calculated with gradient flow agrees with the results from spectral projectors (obtained, though, by direct numerical evaluation of eigenmodes, not stochastically estimated),
with the latter actually much less contaminated by cut-off effects.

\section{Topology in the Plasma}

In this Section we present the  results for topology at nonzero temperature, above
the QCD transition, see Figure~\ref{fig:tcvsmq},  from lattice simulations. 
In the pseudo-critical region the results for the topological susceptibility
complement the analysis of the transition of Section~\ref{sec:spectrum}: at higher
temperature strong interactions approach a perturbative regime, which for
topology is described by DIGA instanton models. 
It turns out that the
topological susceptibility  has a very intricate temperature dependence around the transition.  It has an exponent very different from the DIGA predictions  -- we will see that in QCD the
DIGA predictions are approached only above $T = 350$~MeV. 
Clearly these different behaviours highlight the role of different 
topological objects dominating the QCD partition
function at high temperature, and of their interactions \cite{Larsen:2019sdi,Larsen:2018crg,Larsen:2018xrn}.
We will then contrast the lattice results  with the predictions
for $F(\theta,T)$ from the DIGA and high temperature perturbation theory~\eqref{eq:DIGAF}.
Besides giving information on the nature of the QCD vacuum, this parametrization,
if valid, allows the extrapolation to larger temperatures. 

\subsection{Only gluons: the Yang-Mills theory}

Due to its relative simplicity,
a theory with only gluons  is used
as a laboratory for new ideas. Moreover, stable and controlled
results in the quenched model are an important proof of principle
of the feasibility of calculations in the realistic case. 

The first pioneering studies of finite topology in Yang-Mills theory
on the lattice\cite{Alles:1996nm} were based on the gluonic definition.
One of the first, if not the first,  large scale
analysis of the topological susceptibility at high temperature based
on the index theorem appeared a few years later\cite{Gattringer:2002mr}.
The results covered the range $0.8 T_c < T < 1.5 T_c$.  

Berkowitz, Buchoff, and Rinaldi~\cite{Berkowitz:2015aua} were the first ones to 
implement the suggestion of Ref.~\cite{Wantz:2009it} to
use lattice results as a quantitative input to axion cosmology. 
This motivated a study in an extended range of temperatures. 
They used the gluonic definition for the topological charge with  suitable cooling  
to obtain results up to $T \simeq 700$~MeV. 
Smoothing procedures and multiple volumes were used to control
systematic errors, although a clear continuum extrapolation
was not performed. The results were fitted to the DIGA
motivated  power law decay
$\chi(T) = A\, T ^ b$, with a  quoted value of the exponent
$ b= -5.64(4)$.

In Ref.~\cite{Borsanyi:2015cka} the topological susceptibility
was measured in the even larger
temperature interval $0.9 < T/T_c < 4$,
for several lattice spacing, thus allowing a
controlled continuum extrapolation.
The quoted  result by \cite{Borsanyi:2015cka} is 
$\chi(T) = 0.11(2)(1)(T/T_0)^b$,
$T_0 = 1.02(5)(2)$
($T_0$ should be 1 in unit of $T_c$), $b=-7.1(4)(2)$.
The exponent  is in  nice agreement with the DIGA result, 
whereas the overall normalization of the DIGA prediction still 
differs from the lattice results by a factor of order ten. 
The same group confirmed these results in a later study\cite{Borsanyi:2016ksw}.

Two more recent studies focused on selected large temperatures,
implementing some methodological refinements: master-field simulations 
of the $SU(3)$ gauge theory, leading to high accuracy estimates
\cite{Giusti:2018cmp}, and  improved reweighting 
\cite{Jahn:2020oqf,Jahn:2018dtn}, allowing estimated for temperatures
as high as  $7\,T_c$. 

\begin{figure}
\includegraphics[width=\linewidth]{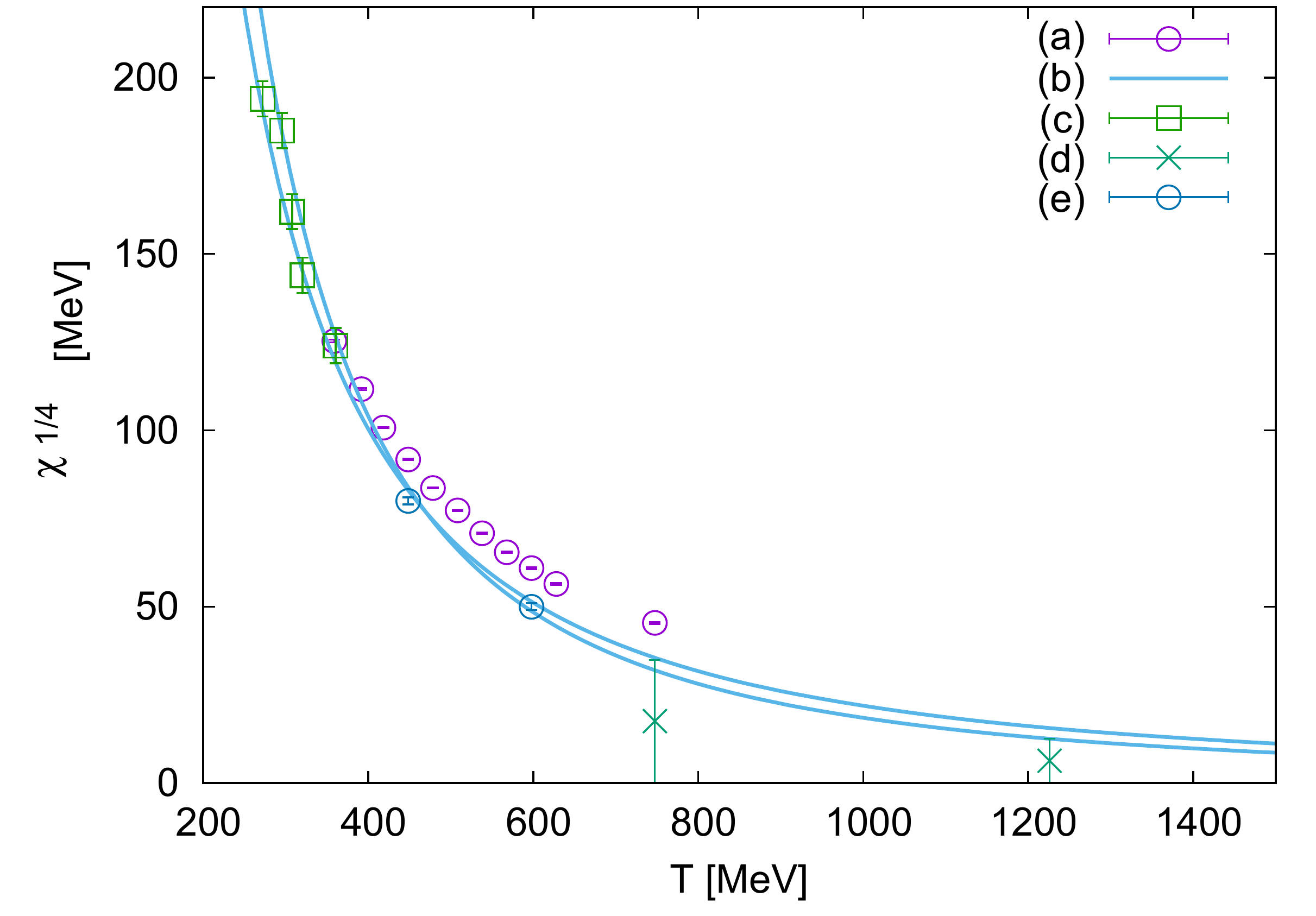}
 \vskip 1 truecm
\caption{The fourth root of topological susceptibility versus the temperature for
the Yang-Mills 
studies reviewed here. Only a representative subset of the results
are shown. (a) is from Ref. \cite{Berkowitz:2015aua}, Table~I. (b) uses
the final fits reported in \cite{Borsanyi:2015cka}. 
 (c) is the earlier results  of \cite{Gattringer:2002mr}, (d) the
results from  improved reweighting \cite{Jahn:2020oqf} 
and  (e) from the master fields simulations \cite{Giusti:2018cmp},
see text for details.}
\label{fig:topsusc-quenched}
\end{figure}

In Figure~\ref{fig:topsusc-quenched} we superimpose the 
numerical results~\cite{Berkowitz:2015aua} with the central values of the fit~\cite{Borsanyi:2015cka}. In the same plot we show 
the 2002 results~\cite{Gattringer:2002mr} and the more 
recent ones~\cite{Giusti:2018cmp,Jahn:2020oqf} 
(omitting the larger temperature in the latter case), 
obtained by use of master field simulations and improved reweighting. 
Interestingly, there is a good agreement up to $T/T_c \simeq 1.5$. 
The fit\cite{Berkowitz:2015aua}, available in the limited temperature range,
gives $b = -5.64(4)$~-- rather stable with respect to changing the interval. 
In~\cite{Borsanyi:2015cka} the quoted result  is $b=-7.1(4)(2)$, where
the systematic error comes from  fits with different initial points
$T/T_c [1.3-1.7]$.  We have checked 
that the results\cite{Berkowitz:2015aua} appear
to be stable with sliding the fit
interval, hence 
a plausible source for the (small) discrepancy may come from residual
finite spacing. Indeed, either Refs.  \cite{Borsanyi:2015cka}  and
\cite{Jahn:2020oqf}  consistently find a decreasing
trend with extrapolations, with errors which may be of the order of several 
percent.  

From the point of view of lattice results, the emerging scenario is rather
pleasing: different groups using different techniques find a nice agreement
once the continuum extrapolation is taken into account. The
continuum extrapolation  seems more critical, 
in agreement with general discussions on lattice topology, at higher
temperatures, while around $T_c$ finite spacing effects seem to be less
severe. Master-field simulations afford an unprecedented
accuracy for topological calculations, however the authors
themselves note that at 
temperatures higher than the ones considered here, 
master-field simulations
of topological susceptibility must be expected to be increasingly 
sensitive to lattice spacing\cite{Giusti:2018cmp}.  
It would be very nice to have results at even higher temperatures, though,
in order to probe more convincingly the approach to the DIGA limit,
which remains a subtle issue. 

Let us stress,
as noticed\cite{Borsanyi:2015cka}, 
that the strong coupling constant, at the relevant
thermal scale of $2\pi T $ is still sizeable
in the considered range of temperatures. 
If we monitor the approach
to DIGA via the exponenent value, we may conclude that the results are rather close to this regime: the fit estimate \cite{Borsanyi:2015cka} is indeed $b=- 7.1(4)(2)$. The approach to the DIGA
result  appears even faster than RG prediction\cite{Borsanyi:2015cka}:
an improved estimate of $b$ shows a mild consistent increase
in the range $ 1.5 < T/T_c < 5$, and the DIGA exponent appears to be 
approached  from below. Looking at the data, we note that close to $T_c$
we have instead a rather fast drop of the topological susceptibility,
perhaps reflecting the fast transition from the
confined to the deconfined phase at the $SU(3)$ (weak)
first order transition. Putting all together, we may expect
 a fast decrease of the apparent
$b$ exponent immediately above $T_c$, then a mild increase 
approaching the DIGA limit. If these two behaviours are smoothly connected,
probably it is not surprising to observe a  proximity to the DIGA
results also at moderate temperatures -- which although may be merely accidental. 

When looking at the absolute value of the susceptibility, one clearly
finds a large deviation from the DIGA prediction, 
by about one order of magnitude\cite{Borsanyi:2015cka}. 
While such large deviations are not uncommon in perturbative QCD,
they indicate that the DIGA limit has not been completely reached yet.
However, Ref. \cite{Borsanyi:2015cka}
have also studied higher order terms in the potential,
analyzing the coefficient $b_2$ in the expansion, which appears
to approach the DIGA limit, even faster than the susceptibility. 

\subsection{Quark-Gluon Plasma}

In recent years, thanks to the methodological progress reviewed above, 
 together with more adequate computer resources,  the 
first results on topology at high
temperature in QCD -- QCD with dynamical
fermions, in lattice jargon -- 
have appeared~\cite{Petreczky:2016vrs, Bonati:2015vqz, Trunin:2015yda, 
Taniguchi:2016tjc, Burger:2017xkz,Burger:2018fvb}.

Although these studies exhibit some common features, a quantitative agreement is still missing.
 Let us stress  that there is a general
consensus that 
 these results have not reached yet the maturity of 
quenched studies, and that there is  room for improvement of the
current understanding. For instance, earlier apparent discrepancies
\cite{Bonati:2015vqz, Trunin:2015yda}   have
been recently successfully resolved \cite{Bonati:2018blm}.

 Particularly significant 
is the onset of the DIGA behaviour.
From the point of axion physics there is a specific interest, besides  probing of the character of the QGP:  if a simple parametrization holds true, 
results could be safely extrapolated to temperatures
$T=\mathcal O(1)$~GeV of cosmological relevance. 

In \cite{Petreczky:2016vrs} the authors use HISQ fermions with 2+1 flavors
exploring the range of temperatures 150~MeV till 800~MeV. 
They use two methods,
gluonic definition with gradient flow and the fermionic definition~\eqref{eq:chit-pbp}.
The results are continuum extrapolated and cross-checked between
different definitions. There is a nice consistence 
between the two methods till  $T < 450$ MeV,
then the results lose significance.
The results are fitted with power laws in two different regions: from
$T < 240$ MeV   the power exponents $b \simeq 6$, while for larger temperatures
the exponent gets closer to the DIGA result. 

A similar interesting cross-check between gluonic and fermionic 
definitions 
was performed  with   the Iwasaki gauge action and 2+1 flavors of 
nonperturbatively $O(a)$-improved 
Wilson quarks,  with temperature ranging from  $T \simeq 174$~MeV  to 
$T \simeq 697$~MeV (the latter on coarser lattices)\cite{Taniguchi:2016tjc}. 
The pion to the rho mass ratio was $0.67$, i.e. heavier than physical. 
The simulations were performed on a single fine lattice, so not
allowing for a continuum extrapolation. 
Nonetheless, the agreement between the two measurements
was very nice, which may be taken as 
an indication of a good control over lattice
artifacts.

An impressive study\cite{Borsanyi:2016ksw} 
covered a range of temperatures up to $T = 2$~GeV, implementing
a number of improvements which allowed a controlled
extrapolations to continuum limit for physical quark masses.
The authors accounted
for the different quark thresholds at finite temperature 
\cite{Laine:2016hma,Laine:2006cp}
by using 2+1 and 2+1+1 flavors, with  physical 
quark masses 
(in the isospin limit, analytically corrected for isospin effects). 
Up to 250 MeV the simulations used 2+1 flavours of dynamical quarks,
either staggered and overlap, 
then 2+1+1 flavors of staggered fermions.
Further on  the step-scaling method for the equation of state  developed by the same group \cite{Borsanyi:2012ve} allowed the extension of the results up to $T \simeq 2$~GeV. 
Degrees of freedom of the Standard Model have been
included as well, leading to an EoS  all the way to the GeV scale. Slightly anticipating the disussion of the final Section, this extended range allowed a controlled estimate
of the number of effective degrees of freedom, hence
of the Hubble constant. 
The results for $N_f =  2+1$ at lower temperatures, and the $N_f = 2+1+1$ results for
$T \gtrsim$  250 MeV can be connected  smoothly.
The topological susceptibility was measured with the gluonic definition,
with suitable smoothing. The extrapolation to the continuum limit was carefully implemented.
The power law decay of the topological susceptibility was monitored,
and found to approach 
the DIGA result $b=8.16$ above  temperatures of about 1 GeV.

The topological susceptibility
was measured with the fermionic method described above
with 2+1+1 flavors
of Wilson fermions at maximal twist, with physical charm and strange mass,
in the temperature range $150\lesssim T \lesssim 500$ MeV
\cite{Burger:2017xkz,Burger:2018fvb}.
The light doublet was degenerate in mass, with pion 
masses of $470$, 370, 260 and $210$~MeV. Since the results 
have not been obtained
for a physical pion mass, an extrapolation was needed:
from the analyticity
of the chiral condensate in the chiral limit above $T_c$ one infers that 
the total susceptibility
is an even series in the quark mass.
Barring unexpected cancellation we may assume that the same holds for the connected and disconnected susceptibilities separately, hence
\begin{equation}
\chi_{top} =  m_l^2 \chi^{disc}_{\pbp} = \sum_{n=0} a_n m_\pi^{4(n+1)}.
\end{equation}
At leading order this coincides with the predictions from DIGA with $N_f=2$, $\chi_{top} \propto m_\pi^4$ -- basically, this implies 
that the disconnected chiral
susceptibility does not depend on the pion mass in the mass range considered.
The results obtained with different lattice spacings showed little or no
residual spacing dependence. A continuum extrapolation was anyway 
performed as well, for the pion mass of $370$~MeV. 

The issue of the continuum limit, once more, is crucial. As anticipated
at the beginning of this subsection,  a detailed analysis of the
continuum extrapolation with the gluonic operator and improved techniques,
focusing on two significant temperatures\cite{Bonati:2018blm} 
resolved an early apparent discrepancy. This is very important
given the large lattice artifacts of the gluonic operator with staggered
fermions \cite{Bonati:2015vqz, Trunin:2015yda}.
The very small decay with temperature of the topological susceptibility
reported in these studies is probably an artifact of 
lattice discretization. We refer to
the cited paper\cite{Bonati:2018blm} for a full discussion.

\begin{figure}
\includegraphics{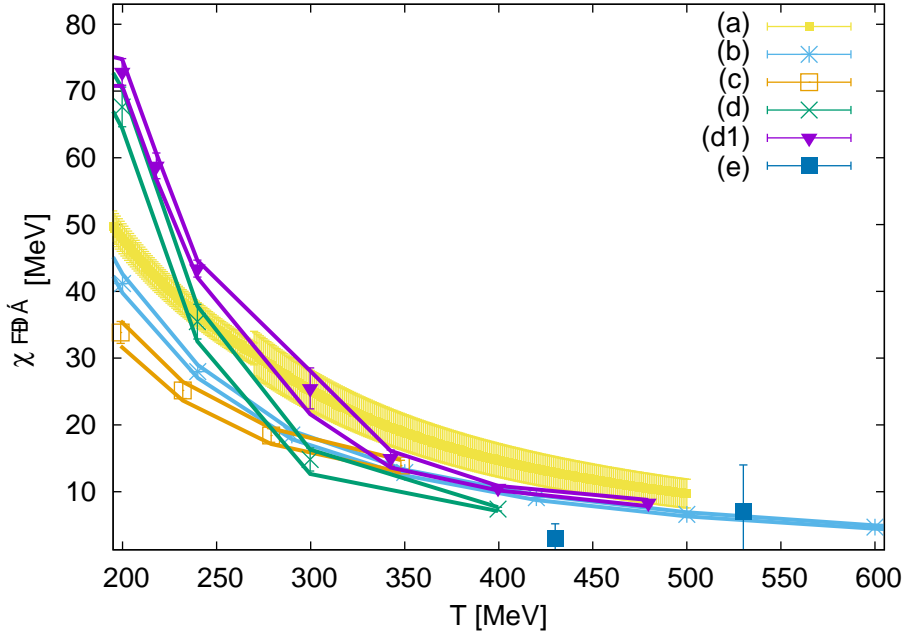}
 \vskip 1 truecm
\caption{
The fourth root of topological susceptibility versus the temperature for
the full QCD studies reviewed here. 
This is only a subset of the results presented in the
papers. (a) shows the gluonic results from Ref. \cite{Petreczky:2016vrs}.
(b) shows the tabulated results from Ref. \cite{Borsanyi:2016ksw}. (c)
Ref. \cite{Taniguchi:2016tjc}: from these results
we may infer the independence on the lattice spacing from the
concordance between gluonic and fermionic methods; moreover we have
rescaled the results with the pion mass. (d) and (d1) shows the
results from Ref. \cite{Burger:2018fvb}, obtained by rescaling from
the two lightest masses, $m_\pi = 220$, 260~MeV. (e) the results
from  Ref.~\cite{Bonati:2018blm}, where a careful
continuum extrapolation with a conservative error estimate was performed.}
\label{fig:topsusc}
\end{figure}

We summarize the results for the topological susceptibility in Figure \ref{fig:topsusc}. 

\begin{figure}
\includegraphics[width=0.9\textwidth]{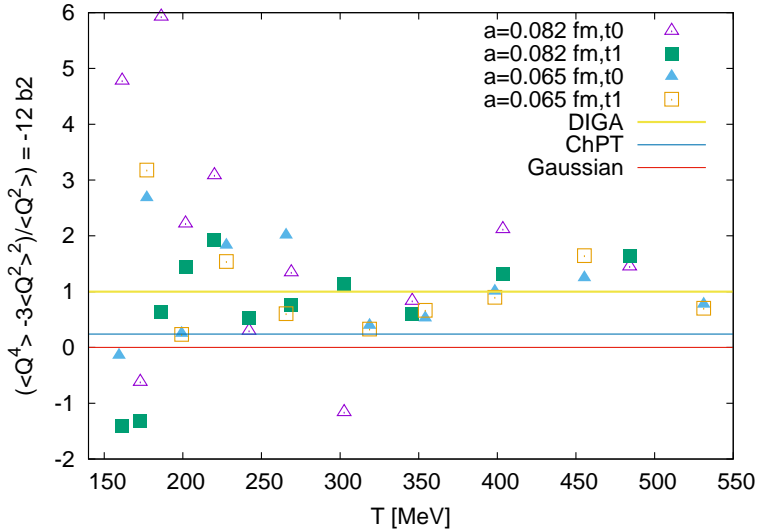}
\caption{$b_2$ from lattice simulations  with $N_f = 2+1+1$ ,
and lattice spacings as indicated. $t_0$ and $t_1$ correspond to the two different levels of cooling.  The gauge configurations are the same as in \cite{Burger:2018fvb}. }
\label{fig:b2}
\end{figure}

Beyond the topological susceptibility, the potential
has been investigated up to the cumulant $b_2$.
Its behaviour supports the approach to DIGA
\cite{Bonati:2015vqz,talktrento} already at $T = 400$ MeV, supporting
the conclusion that the DIGA behaviour is reached already at
these temperature, see Figure \ref{fig:b2}, and the discussions
in\cite{Bonati:2015vqz}. One has to underscore that $b_2$ merely tests the cosine
form of the potential: a simple cosine tells us that
the vacuum is dominated by instantons of 
positive and negative unit charge. This
does not have any direct implications on the amplitude 
of the topological susceptibility, and there is no  contradiction
between the cosine shape of the potential,
and the amplitude of the topological susceptibility. 

There is  an interesting difference between the Yang-Mills and the quenched
results, which is highlighted and discussed\cite{Bonati:2015vqz}: 
in Yang-Mills the approach to the DIGA value of the exponent is
faster, and the DIGA result is approached from below. In QCD, the approach
is slower and happens from above. These different behaviours may shed light
on the different instanton dynamics\cite{Bonati:2015vqz,Larsen:2019sdi,Larsen:2018crg,Larsen:2018xrn}. 
On a more technical note, and interestingly, $b_2$
seems less sensitive to the spacing effects with respect to the topological
susceptibility. In Figure~\ref{fig:b2} we show 
the results on $b_2$ obtained on the same gauge configurations \cite{Burger:2018fvb}, but with 
the gluonic method, for different lattice spacings and different level of cooling. 
The results -- within the large uncertainties -- are consistent 
with Ref.~\cite{Bonati:2015vqz}. 

\begin{figure}
\centering
{\includegraphics[width=0.65\textwidth]{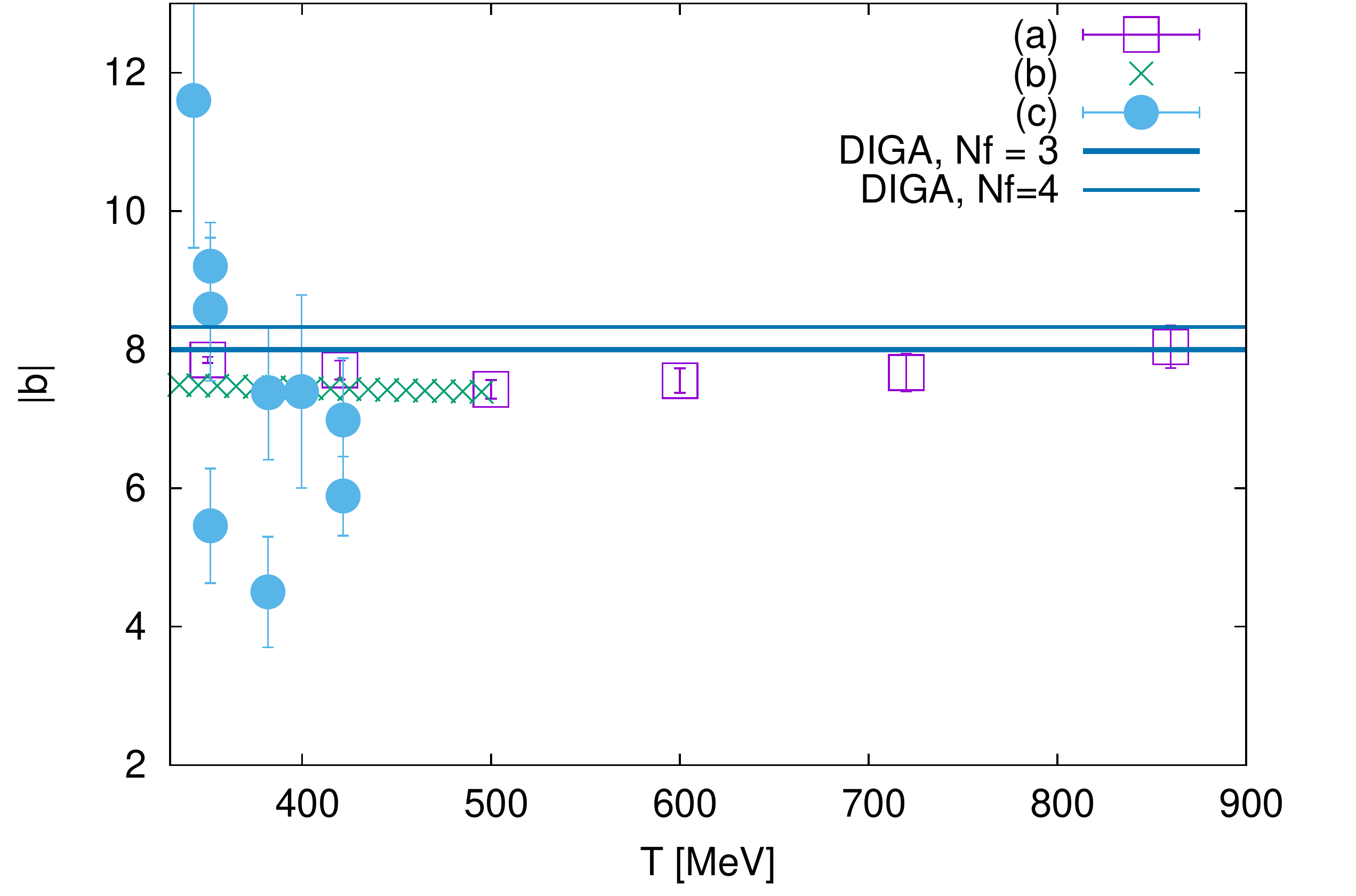}
\includegraphics[width=0.32\textwidth]{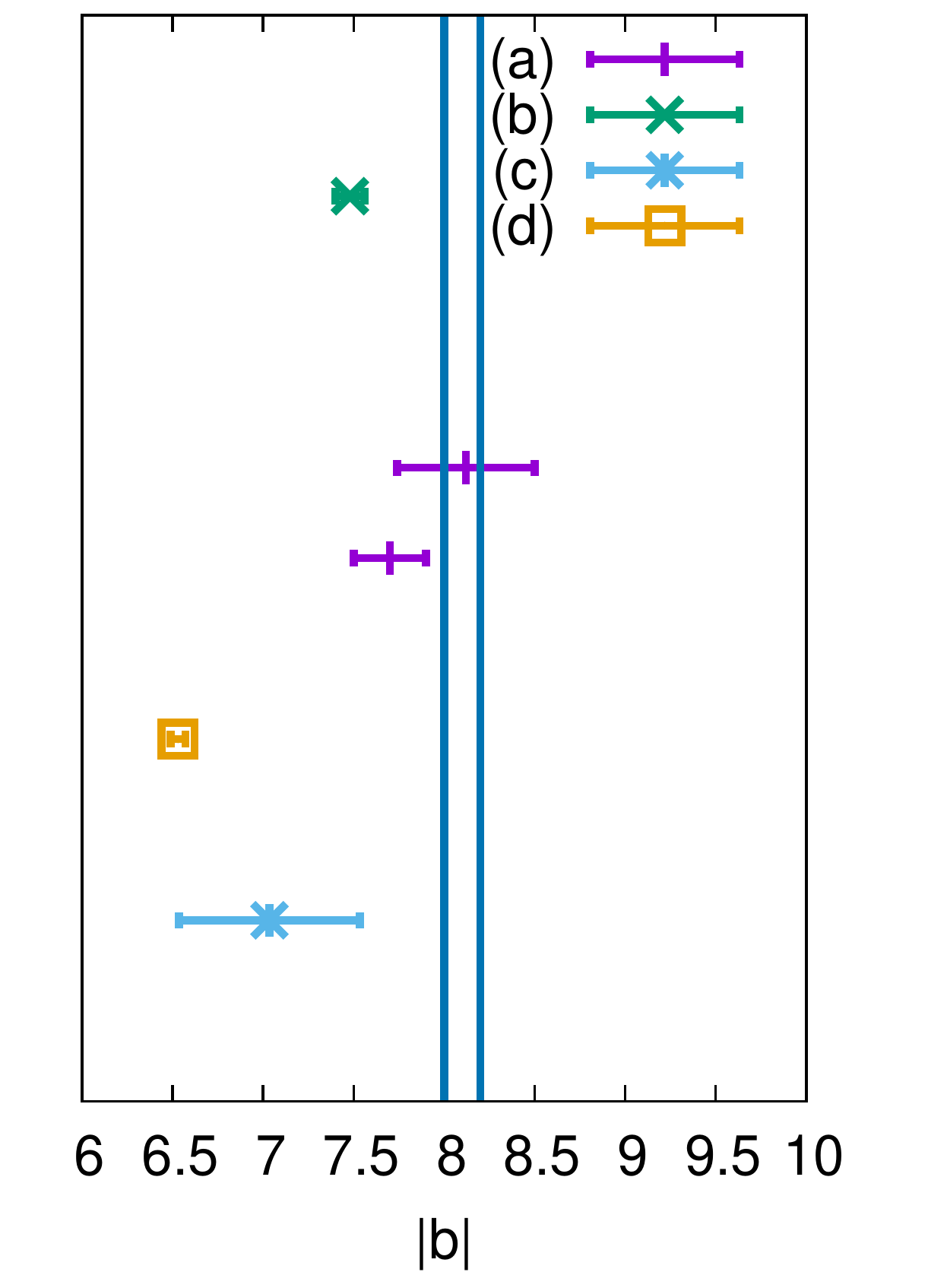}}
\caption{The absolute value of the exponent characterizing the power-law decay of the topological susceptibility at high temperatures: in the left hand side plot we show (a) log-derivative of the tabulated continuum results 
of $\chi_{top}$~\cite{Borsanyi:2016ksw}; (b)  log-derivative of the continuum fit to  the data of Ref. \cite{Petreczky:2016vrs}; (c) log-derivative  from the  two lowest pion masses from the Wilson twisted study\cite{Burger:2018fvb}.  The results are contrasted with the DIGA limit prediction; in the right hand side plot the same results are summarized: for (a) we plot the results at $T = 420, 2100$ MeV, for (b) the average effective exponent in the temperature range considered; (c) shows the average and dispersion of the fit results for different pion masses in the temperature range $[200:600]$ MeV; (d) is result of the fit to the data of Ref.\cite{Taniguchi:2016tjc} in a restricted temperature range $T < 350$ MeV. }
\label{fig:effexp}
\end{figure}

As we have briefly reviewed, all the authors contrast their power-law
fits with the DIGA prediction, finding 
a reasonable agreement for $T \gtrsim 350$ MeV.
To appreciate the approach to DIGA in some more
detail, 
we present a summary in Figure~\ref{fig:effexp},
where we have considered  the 
logarithmic derivative of
the topological susceptibility \cite{Borsanyi:2016ksw,Burger:2018fvb}.
The log derivative\cite{Burger:2018fvb}
is rather noisy and the statistical errors
overcome any mass/spacing systematics.  

The average result confirms the 
approach to the DIGA limit  above $T \sim 350$~MeV
For
Ref.\cite{Borsanyi:2016ksw} the log-derivative was computed using their
tabulated results for the continuum topological susceptibility. 

In comparison with the Yang-Mills results, the errors on the topological susceptibility are much larger. 
However, there are interesting common trends: 
at rather high temperatures, all the studies
confirm a strong correction factor, 
despite the fact that the
exponent is close to the DIGA prediction, 
a trend which was already observed in pure gauge. 

Closer to $T_c$ we observe the largest spread  in the results: 
perhaps it is not too surprising that a difficult observables feel
the intricacy of the pseudocritical region~-- investigating
and reaching a final conclusion on the behaviour of topology
has bearing on the general properties of the transitions reviewed in Section~\ref{sec:spectrum}. 

\section{The QCD Axion }
\label{sec:axion}

We have mentioned that within the
predicted bound $f_A \gtrsim 4 \times 10^8$ the axion may contribute
to contemporary density of cold dark matter. How this is realised, and the quantitative significance of the contribution, 
depends on the axion's 
cosmological history~\cite{Marsh:2015xka,Sikivie:2006ni,Ringwald:2012hr}. 
Early studies using models\cite{Wantz:2009it,Turner:1985si}
identified the relevant temperature range $\mathcal O(1)$ GeV. A recent estimate of the temperatures of
relevance is 
$500\lesssim T \lesssim 1500$ MeV \cite{Moore:2017ond}.
To be useful for cosmology lattice
simulations need to produce controlled results in this range. 

In a nutshell, there are two main sources of axion production: by a thermal
bath, producing hot axions, and by the so-called realignment
mechanisms, which produces cold axions
\cite{Preskill:1982cy,Abbott:1982af,Dine:1982ah}. These cold axions
are those that  can provide observable dark matter.
We will briefly review here
how the typical parameters of axion's evolution~-- in particular, the
temperature dependence of the topological susceptibility -- 
affect the final density of dark matter. 

The cosmological history of the axion begins with the PQ symmetry breaking -- at $T_{PQ} \simeq f_A$.
At this point 
the axion field $a(x)$ will set somewhere at the bottom of the Mexican
Hat, with an angle $\theta_1$  "misaligned"
with the conserving CP minimum $\theta = 0$ of the 
contemporary potential.  Axion is then a massless
Goldstone boson. As time passes,  temperature goes
further down, and the dynamics becomes sensitive
to the $U(1)$ breaking term (the topological fluctuations),
i.e. to the axion potential $V(a) = \chi_{top}(T)(1 - \cos (a/f_A))$.
The axion is now a massive pseudo-Goldstone boson, "oscillating"
around the minimum\footnote{Actually, $U(1)$ is broken to a discrete subgroup,
which may produce domain walls -- for simplicity, we have ignored this possibility
in the discussion of QCD symmetries in Section~\ref{sec:spectrum}.}.

The axion contribution to the cosmological density could
be, in principle, estimated by solving the equation of motion
for the axion field in spacetime. We give here a brief
summing up of the main approximations, which have been
used to arrive at a simple expression relating contemporary
axion mass with the parameters characterizing the topological
susceptibility evolution with temperature. In the following
we will contrast the lattice data with the DIGA inspired  behaviour, Eq.~\eqref{eq:DIGAF},
\begin{equation}
\chi_{top} = A\,T^b
\label{eq:topc}
\end{equation}
and we will discuss the impact on the results for the bounds on the
contemporary axion mass inferred from the variability
of the parameters $A$ and $b$.

A few years ago\cite{Berkowitz:2015aua} the first lattice
studies aimed at extracting limits
on the axion mass appeared, initially in the quenched limit, followed by simulations in full QCD\cite{Borsanyi:2016ksw, Bonati:2016tvi,Petreczky:2016vrs,Burger:2017xkz,Burger:2018fvb}. In the previous
Section we have reviewed the results for the topological
susceptibility, and here we will discuss their impact on
axion's density and mass. 

On general grounds, the results will depend on the initial
angle $\theta_1$:
if the PQ transition happened during, or before,  the inflation,
the initial angle $\theta_1$ will be made homogeneous in all the
space-time. $\theta_1$  is arbitrary, 
the final predictions will depend on it, leading to  large uncertainties.
After inflation, contributions from different
regions of space time will be averaged, and there will be no dependence
on the initial angle. It is customary to refer to these two scenarios
as pre-inflationary and post-inflationary, respectively.

\begin{figure}
\includegraphics[width=\textwidth]{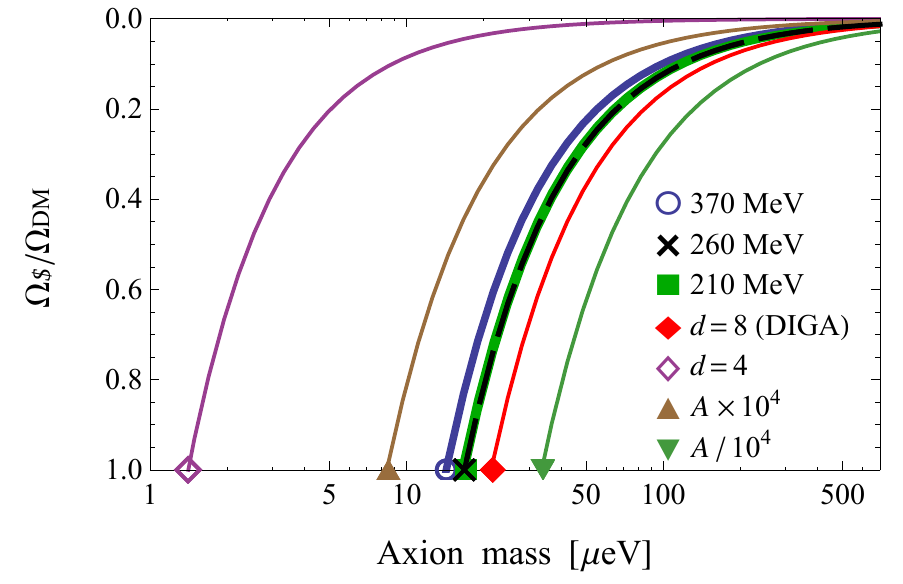}
\caption{The axion contribution to Dark Matter versus the axion mass\cite{Burger:2018fvb}: the first three lines
show the results for three pion masses,
$m_\pi = 370$, 260, and 210 MeV, respectively,  
all rescaled to the physical pion mass according to $\chi_{top} \propto m_\pi^4$.
The other lines are mock data meant to study the sensitivity on the errors to
the fit parameters of the topological susceptibility $\chi_{top} \simeq A\, T^{-d}$,
see text for discussions. Figure taken from Ref.\cite{Burger:2018fvb}.}
\label{Fig:axions-fraction}
\end{figure}

The topological susceptibility enters 
the equation of motion for the axion degree of freedom $\theta$
via the potential  
 $V(\theta) = \chi_{top}(\theta) f(\theta)  \simeq   \chi_{top}(\theta)  (1 - \cos \theta)  $,
\begin{equation}
    \ddot\theta + 3H \dot\theta + V'(\theta) = 0,
\label{eq:evolv}    
\end{equation}
where $H$ is the Hubble parameter\cite{Preskill:1982cy,Wantz:2009it}.
At early times the solution of Eq.~\eqref{eq:evolv} is a constant $\theta = \theta_1 = \text{const}$.
When the axion mass is
of the order of the inverse of the Hubble parameter:
$3 H(T) = m_A(T) = \sqrt{\chi_{top}(T)}/f_A$, 
 $\theta$ starts oscillating. 
The Hubble parameter can be approximated as
\begin{equation}
   H(T) = \frac {\pi g_{*}^{1/2}(T)\,T^2}{\sqrt{90} \,M_{Pl}}.
\end{equation}
$g_{*}(T)$ is the effective number of relativistic degrees of freedom,
which has been computed up to $T=3$~GeV in the study reviewed above\cite{Borsanyi:2016ksw}.
To obtain simple expressions in closed form we employ the power-law parameterization
$g^*(T) = 50.8 \,(T/(\text{MeV}))^{0.053}$
reproducing the results\cite{Borsanyi:2016ksw} 
up to a few percents in the temperature interval $800< T <1500$~MeV.

The knowledge of the temperature dependence of the topological
susceptibility determines the time of beginning of the
oscillations as a function of $f_A$, or, equivalently,
of $m_A$, if one trades $f_A$ for the zero temperature topological susceptibility
and the axion mass via $\chi_{top} = f_A^2 m_A^2$.
Since then, time-averaged oscillations behave as
pressureless dark matter \cite{Preskill:1982cy}, and
the energy density $\rho_A(T)$ of the oscillating axion field is 
approximately the same as a collection of axions
at rest:  $\rho_A(T) \approx 1/2 m_A^2(T) f_A^2 \theta^2$. Hence the number
density  $n_A (T) = \rho_A(T)/m_A(T)$ can be estimated
as $n_A(T) \approx 1/2 m_A(T) f_A^2 \theta^2$ .

The axion-to-entropy ratio remains constant after the beginning of the oscillations:
knowing this ratio at the beginning of the oscillation,
one can estimate the contribution of the axion to today's
density  $\rho_{A,0} = \dfrac{n_A(T)}{s(T)} m_A s_0$,
where $s$, $s_0$ are the entropies at time $T$,
and of today, and $\Omega_A = \dfrac {\rho_{A,0}}{\rho_c}$,
with $\rho_c$ the critical density.  

In this simplified picture, the resulting axion density  depends 
on the axion mass, on the topological susceptibility parameters $A,b$
and  on $\theta_1$, the initial misalignment
angle. For instance\cite{Burger:2018fvb}: 
\begin{equation}
\rho_A(m_A) = \rho_A(m_A,b,\theta_1)=
\rho_A(m_A) = C(A,b,\theta_1) m_A^{-\frac{3.053 -  b/2}{2.027 - b/2}},
\label{eq:dens}
\end{equation}
where we have highlighted the most relevant dependence on the exponent $b$. 
The coefficient $C$ depends on
$b$ only weakly, and of course includes the relevant
cosmological constants. 

Note again the dependence on $\theta_1$: in a post-inflactionary  scenario
the misalignment angle can be averaged over. Anharmonicity effects
on the potential may be included as well, 
but have little impact\cite{Berkowitz:2015aua}. 
The pre-inflactionary scenario allows more freedom:
clearly one
can cover a wide range of axion masses by suitably adjusting the initial misalignment
angle $\theta_1$\cite{Ringwald:2018xlf}.

Considering now the post-inflactionary scenario, 
the final density of axions can be contrasted with 
the known dark matter density to arrive
at a bound on the contemporary
axion mass\cite{Preskill:1982cy,Abbott:1982af,Dine:1982ah}.
However, one has to deal with a more complicated dynamics
associated, essentially, with the possible presence of strings,
domain walls  and other forms of non-homogeneous structures such as archioles\cite{Ringwald:2018xlf,Klaer:2017ond,Saurabh:2020pqe,Sakharov:1994id,Sakharov:1996xg}.
This limits the contribution to dark matter coming from axions alone, 
and it is a subject of contemporary research \cite{Jahn:2020oqf,Jahn:2018dtn}:
even in the unlikely case that axion dynamics is the only responsible
for dark matter, one cannot reach the overclosure bound with axion density alone.  

In short, and simplifying, 
the amount of post-inflactionary axion dark matter can then be divided into a re-alignment
contribution and a contribution from axionic strings and domain walls.

Consider now the contribution from re-alignment. 
The estimate \cite{Borsanyi:2016ksw} gives an absolute lower bound of 
$m_A =28(2)$~$\mu$eV, and $m_A = 50(4)$~$\mu$eV if the contribution from misalignment
mechanism contributes 50\% to dark matter.  

The study\cite{Petreczky:2016vrs}
combines lattice results and DIGA predictions and
reports $f_A \le 1.2 (0.152) \times 10^{12}$~GeV,
when this contribution
matches the total amount of Dark Matter. 

The results \cite{Burger:2018fvb} are presented in a
graphic form in Figure~\ref{Fig:axions-fraction}: 
the axion's fractional contribution to Dark Matter is shown 
versus the axion mass for various situations.
The first three lines are obtained from the results of three different pion masses, 
rescaled to the physical pion mass.
The discrepancy between the results are very 
small at practical level. The impact of the uncertainties
may be estimated by plotting a few mock curves, varying the parameters $A$ and $b$
using the central results as baseline. 
In all cases  the intercept with the abscissa axis (overclosure bound) defines the absolute lower
bound for the axion mass \cite{Burger:2018fvb} $m_A \simeq 20(5)$~$\mu$eV, where the error is estimated
from the spread of the results of the (plausible) fits.  Clearly the bounds are robust against 'small' changes of parameters.  However, a significant 
variability  remains, and, in particular, a slower decay would significantly lower the limit.

Even  larger uncertainties come from the strings and domain
wall contributions~-- recent estimates\cite{Ballesteros:2019tvf,Ballesteros:2016xej}, using as an input the results\cite{Borsanyi:2016ksw} for
the topological susceptibility, conclude that 
the total axion dark matter, computed
adding the contributions  from strings and  domain walls to the one from misalignment,
fits the total amount of dark matter if $m_A > 50$~$\mu$eV. Recent researches including 
an improved analysis of the string dynamics\cite{Klaer:2017ond,Gorghetto:2018ocs,Klaer:2019fxc},
give
$m_A > (26.2 \pm 3.4)$~$\mu$eV~\cite{Klaer:2017ond}.

As for the pre-inflactionary scenario, the freedom of adjusting $\theta_1$ leads to a great variability for
the axion density (see Eq.~\eqref{eq:evolv}), 
and the final result is less constrained, with values of the 
lower bound for the axion mass which may be reduced  by two orders of magnitude. 

\begin{figure}
    \centering
    \includegraphics[width=0.9 \textwidth] {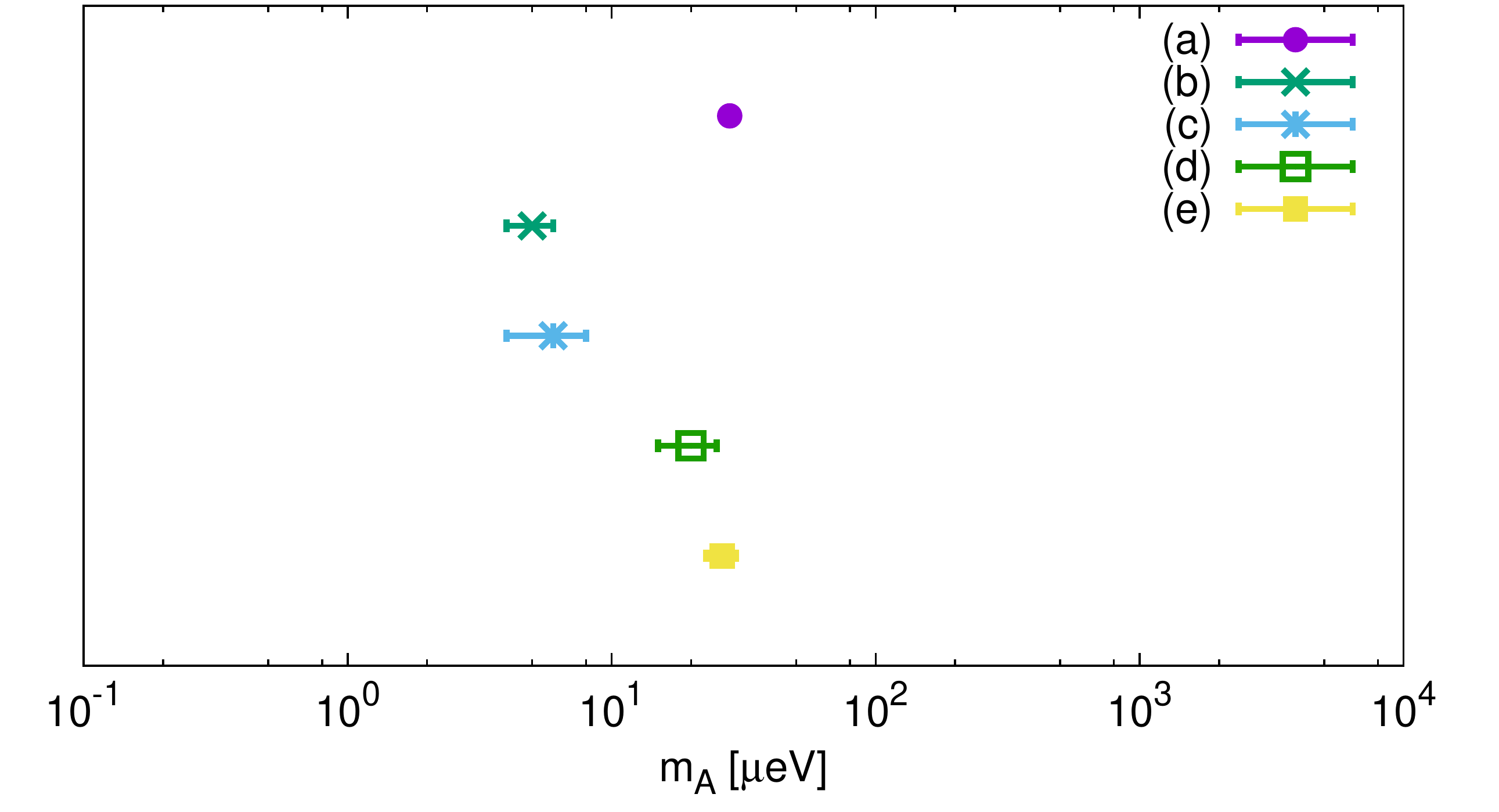}
    \caption{The results for the lower bound on the post-inflactionary axion mass discussed in the text. The results in (a)\cite{Borsanyi:2016ksw}; (b)\cite{Petreczky:2016vrs} (c)\cite{Bonati:2015vqz}  (d)\cite{Burger:2018fvb} are the contributions from misalignment from lattice simulations; (e)\cite{Klaer:2017ond} includes contributions from axion string dynamics. The $x$ axis approximatively spans the allowed range from cold axions: beyond the upper limit the contibution from axions would become negligeable. In the preinflactionary scenario the variability of the $\theta$ angle allows the smaller values of the axion mass corresponding to the lower limit set by astrophysical constraints.}
\label{fig:ax}
\end{figure}

Figure~\ref{fig:ax} summarizes the results for the acceptable range of axion masses
emerging from these discussions. 

The lower bounds on the post-inflactionary
axion mass from the lattice calculations,
in the case of QCD axions saturating all the Dark Matter, are marked as
(a-d). In very short summary (a)\cite{Borsanyi:2016ksw} relies on  a convincing continuum
result, and on direct measurements in a broad range of temperatures; (b)\cite{Petreczky:2016vrs} and 
(d)\cite{Burger:2018fvb,Burger:2017xkz} 
explored a more restricted temperature range, and relied on extrapolations, in addition
(d) needed a rescaling to physical pion mass; (c)\cite{Bonati:2015vqz} suffers from lattice artifacts,
which have been already to some extent clarified\cite{Bonati:2018blm}. Finally, (e)\cite{Klaer:2017ond}
includes contributions from axion string dynamics, still using input from lattice 
results\cite{Borsanyi:2016ksw}.

From Figure \ref{Fig:axions-fraction}, and from similar results on the other works, one can 
see how an increasing axion mass would correspondingly lower the contribution to Dark Matter:
indicatively, one places a tentative upper bound on the axion mass at $1500$~$\mu$eV, when the 
axion dark matter fraction is at the level of one percent.

Finally, in the pre-inflactionary situation, different choices for the initial misalignment
angle basically would allow any value for the axion mass within the entire range spanned by $m_A$ 
in Figure \ref{fig:ax}.

Axion masses in the window discussed will be probed in the upcoming
decade by  direct detection experiments such as MADMAX, ADMX, CAPP,
HAYSTAC, RADES, ORPHEUS and others. Recent reviews discuss these prospectives
at different level of details\cite{Jaeckel:2020dxj,Irastorza:2018dyq}.

\section{Parting remarks}

Axions are well motivated candidates for Dark Matter, and QCD topology in the range $500 \lesssim T \lesssim 1500$ MeV provides important input
for the prediction of some of their properties. 
We have reviewed the current status of studies of topology in  QCD, and we have discussed the implications on axion physics, in particular the
limits on the post-inflactionary axion mass. 

At zero temperature,  QCD topology is a mature field,
and the lattice results compare well with phenomenology
and other estimates from model calculations. At very high temperature topology is constrained by instanton models. The vast range of temperatures
in between is studied by ab-initio 
simulations of QCD on the lattice, which
we have reviewed in this paper. 

In the pure gauge theory,  topology is becoming increasingly accurate: results from different
approaches are in good agreement, within small
residual errors.

The first results for topology in high temperature QCD  are
relatively recent. We have discussed the steady progress and the remaining uncertainties of the lattice calculations.
For temperatures  above 350 MeV  topology faces in principle  several technical problems, which we have reviewed. Nonetheless, a consistent picture emerges, indicating  the approach to the DIGA limit for the temperature dependence of the susceptibility, and to the periodical form of the potential. A complete quantitative agreement among different
lattice results has not been reached yet.
There is however a substantial agreement on the possible sources of not-yet-quantified systematic errors, and improvements are within reach. 

Accurate measurements of the topological susceptibility in QCD are important in order to quantify the mis-alignment contribution to axion dark matter.  Beside mis-alignment,  the current contribution of the axions to Dark Matter is affected by many sources of uncertainty, including those coming from topological defects which necessarily accompany the production of post-inflactionary axions. Lattice results are also used as input to the analysis of these contributions.

The bounds on the axion mass derived from 
lattice results point at a range  which is well within the reach of current experiments. The lattice studies predict a lower bound on  the post-inflactionary axion mass from mis-alignment ranging from 
$m_A \simeq 5$~$\mu$eV till 
$m_A \simeq 28$~$\mu$eV,
assuming that all Dark Matter is made of axions. 
The sensitivity of the axion mass to the fraction of Dark Matter contributed by the mis-alignment mechanism can be read off Figure~\ref{Fig:axions-fraction}.
The analysis of the contribution from  strings, which still
relies on lattice results, gives a lower bound $m_A = (26.2 \pm 3.4)$~$\mu$eV.  Further effects are under active investigation. 

One final comment concerns the region around the QCD phase transition, where  the interrelation
of topology,  chiral and axial symmetry
has a profound influence.  The general picture there is not completely understood, and constitutes a  main focus
of the current research in the physics of strong interactions.

\section*{Acknowledgments}
We have enjoyed discussing these topics with our collaborators and many colleagues: it is a pleasure to thank all of them.  In particular, 
we are grateful to Krzysztof Cichy, Heng-Tong Ding, Roberto Frezzotti, Karl Jansen, 
Alessandro Mirizzi, Guy Moore, Giancarlo Rossi, Sayantan Sharma, 
and Giovanni Villadoro for correspondence and discussions while preparing this review. 
A.T. acknowledges support from the "BASIS" foundation and RFBR grant No. 18-02-01107.
M.P.L. wishes to thank the Bogoliubov Laboratory of Theoretical Physics of JINR, Dubna and
the Galileo Galilei Institute, Firenze for their kind hospitality.
We thank CINECA for computing support under the INFN--CINECA agreement INFN20sim and ISCRA project IsB20PCHSHT. 
This work is partially supported by  STRONG-2020, a 
European Union’s Horizon 2020 research and innovation programme under grant agreement No. 824093,
and the European COST Action
CA15213 "Theory of hot matter and relativistic heavy-ion collisions" (THOR).

\bibliographystyle{ws-ijmpa}
\bibliography{topaxion}
\end{document}